\documentclass[aps,twocolumn,nofootinbib]{revtex4}
\usepackage[latin1]{inputenc}
\usepackage{epsfig}
\newcommand{\beq}{\begin{equation}}
\newcommand{\eeq}{\end{equation}}
\newcommand{\la}{\langle}
\newcommand{\ra}{\rangle}
\usepackage{amsfonts}

\begin{document}

\title{Parametric invariance}

\author{Mário J. de Oliveira}
\affiliation{Universidade de São Paulo, Instituto de Física,
Rua do Matão, 1371, 05508-090 São Paulo, SP, Brazil}

\begin{abstract}

We examine the development of the concept of parametric invariance 
in classical mechanics, quantum mechanics, statistical mechanics,
and thermodynamics, and particularly its relation to entropy.
The parametric invariance was used by Ehrenfest 
as a principle related to the quantization rules of the old 
quantum mechanics. It was also considered by Rayleigh
in the determination of pressure caused by vibration,
and the general approach we follow here is based on his.
Specific calculation of invariants in
classical and quantum mechanics are determined.
The Hertz invariant, which is a volume in phase space,
is extended to the case of a variable
number of particles. We show that the slow parametric
change leads to the adiabatic process,
allowing the definition of entropy as a parametric
invariance. 

\end{abstract}

\maketitle

\section{Introduction}

When a mechanical system is under the influence of 
a disturbance caused by a time variation of one of its
parameters, we expect its properties to
change. However, it was found
that there are some properties that remain invariant
if the parameter changes very slowly.
It is customary to trace the origin of this
type of invariance to the Solvay Congress 
held in Brussels in 1911 \cite{solvay1912}.
During the discussion that followed
the Einstein lecture,
Lorentz remembered a conversation he had with Einstein
sometime earlier. In this conversation, he
asked him how the energy of a 
simple pendulum varies when its lengths is shortened
by holding the string between two fingers and sliding
down. Einstein replied that, if the length of
the pendulum is changed in an infinitely slow manner,
the energy varies in proportion to the frequency of
oscillations. In other words, the ratio between energy
and frequency remains invariant.

This concept of invariance appeared more consistently 
in the writings of Ehrenfest 
\cite{jammer1966,klein1985,navarro2004,navarro2006,perez2009}
on the formulation of the theory of quanta, later
called old quantum theory \cite{jammer1966,terhaar1967,waerden1967}.
He used these invariants to formulate his procedure
for obtaining quantized states. To this end, he introduced
in 1913 \cite{ehrenfest1913b,ehrenfest1913c}
a hypothesis according to which the allowed motions
of a system are transformed into allowed motions
if the system is affected by a reversible adiabatic change.
In a paper published in the following year, Einstein
used the Ehrenfest hypothesis, and called it the
{\it adiabatic hypothesis} \cite{einstein1914}.
The invariant quantities
resulting from a reversible adiabatic change 
Ehrenfest called {\it adiabatic invariants} in a paper
of 1916 \cite{ehrenfest1916,ehrenfest1917,ehrenfest1917a}.
In that same paper Ehrenfest 
explained what he meant by a reversible
adiabatic change: it is an influence on the system in
which the parameters change in an infinitely slow way.

In spite of the explanation given by Ehrenfest,
the influence of the infinitely slow change of a
parameter became associated to the term adiabatic.
Jeans considered the term not particularly
a happy one \cite{jeans1921}.
Accordingly, we find it more appropriate to name
it by its definition and not by its consequences,
and call it a {\it slow parametric action}, which in
addition avoids the reference to any thermodynamic meaning.
The invariant that results from this action we call
parametric invariant. 
We reserve the term adiabatic invariant for a thermodynamic
quantity that is constant along a slow adiabatic process,
an example of which is the well known Poisson relation
between pressure and volume for an ideal gas. 

Here we review the concept of parametric invariance
through the critical analysis of its evolution and
how it is treated in classical mechanics and 
in quantum mechanics as well as its relation
with thermodynamics particularly with entropy.
The parametric invariance is included as a subject  
of classical mechanics
\cite{born1927b,landau1960,arnold1963,terhaar1964,%
arnold1978,goldstein1980,jose1998},
usually connected with the technique of action-angle variables.
It is treated in quantum mechanics 
\cite{messiah1966,gasiorowicz1974,griffiths1994,sakurai2010},
kinetic theory and statistical mechanics
\cite{becker1967,munster1969,toda1983},
dynamics of charged particles
\cite{chandrasekhar1958,gardner1959,northrop1963,lehnert1964},
and has been applied to specific problems by several authors 
\cite{morton1929,bhatnagar1942,parker1971,gignoux1989,%
crawford1990,mohallem2019}, particularly in modern computer
calculations to determine entropy and free energy
by the method of adiabatic switching \cite{watanabe1990,dekoning1996}

\section{Ehrenfest principle}

Ehrenfest enunciated his hypothesis in 1916 in
the following terms
\cite{ehrenfest1916,ehrenfest1917,ehrenfest1917a}:
If a system be affected in a reversible adiabatic way,
allowed motions are transformed into allowed motions.
In the paper of 1914, Einstein stated it in the
following terms \cite{einstein1914}: With reversible
adiabatic changes of a parameter, every quantum-the\-oretically
possible state changes over into another possible state.
In both statement, reversible adiabatic change is to be
understood as a slow variation of a parameter.

In his paper of 1916 \cite{ehrenfest1916,ehrenfest1917},
Ehrenfest considered a periodic
system and showed that the time integral of twice the
kinetic energy $K$ over a period,
\beq
I = \int 2 K dt,
\label{13}
\eeq
is a parametric invariant. Defining the time
average $\bar{K}$ of the kinetic energy by
\beq
\bar{K} = \nu \int K dt,
\label{13a}
\eeq 
where $\nu$ is the frequency, the inverse of the period,
the invariant is equivalent to $2\bar{K}/\nu$.
For a harmonic oscillator the energy $E$ is twice
the kinetic energy and the invariant becomes $E/\nu$.

Ehrenfest had already presented the
invariant (\ref{13}) in his publication of 1913 but now he
provided a demonstration through the use of
the Lagrange analytical theory. Considering that
the kinetic energy $K$ of a system with many degrees of
freedom is a quadratic form in the variables
$\dot{q}_k$, the time derivative of the
coordinates $q_k$, the
Euler theorem on homogeneous functions allows
us to write 
\beq
K = \frac12 \sum_k p_k \dot{q}_k,
\eeq
where $p_k=\partial K/\partial\dot{q_k}$ is the
momentum conjugate to $q_k$. Using this expression
Ehrenfest writes the integral (\ref{13}) in the form
\beq
I = \sum_k \int p_k dq_k.
\label{14}
\eeq
The geometrical interpretation of this expression
was given by Ehrenfest as follows. In the phase
space, the representative point of the system
describes a closed curve which projects closed
curves on each one of the planes $(q_k,p_k)$. 
Each term
\beq
I_k = \int p_k dq_k 
\label{14a}
\eeq
of the sum in (\ref{14}) represents the area of each
one of the projected closed curves.

In 1915, Wilson \cite{wilson1915} and by Sommerfeld
\cite{sommerfeld1915a}, independently postulated the
quantization rule by the use of phase integrals,
\beq
\int p_i dq_i = n_i h,
\label{14b}
\eeq
where $n_i$ is an integer number and $h$ is the Planck
constant. These phase integrals were shown by 
Schwarzshild \cite{schwarzschild1916} and by 
Epstein \cite{epstein1916a,epstein1916b} to
emerge when it is possible to separate variables
by using the Hamilton-Jacobi theory to 
systems in which the variables can be separable.
The Wilson-Sommerfeld rule is then applied
to each pair of these separable canonically conjugate variables,
called action and angle variables by Schwarzschild
\cite{schwarzschild1916}.

At the end of his paper of 1916, Eherenfest asked
himself whether the phase integral (\ref{14b})
appearing in the papers of Schwarzschild and Epstein 
could also be an invariant.
The demonstration that indeed each one
of these phase integrals is an invariant was shown by
Burgers in 1916 \cite{burgers1917,burgers1918}.
According to Burgers,
if the momentum $p_k$ in the integral (\ref{14a}) depends
only on $q_k$ then $I_k$ is an invariant. 
Notice that Ehrenfest had showed that the {\it sum} of integrals
of the type (\ref{14a}) is an invariant, nothing being said
about each one of them.

In 1913 Bohr proposed his atomic model based on the
assumptions that the electron describes stationary orbits
around the nucleus \cite{bohr1913}.
Bohr assumed that the frequency $\nu$ of
the radiation emitted is half the frequency of revolution
of the electron and that the amount of energy emitted is
$h\nu$ times an integer $n$.
From these assumptions he obtained the binding energy $E$
of the electron as
\beq
E = \frac{2\pi^2 m e^4}{h^2 n^2},
\eeq
where $e$ is the charge of the electron and $m$ its mass.
Another fundamental assumption made by Bohr was as follows. When
the electron passes from one stationary orbit to another,
the loss of energy in the form of radiation is equal to $h\nu$. 

As a way of justifying the stationarity of the orbits,
Bohr employed the Ehrenfest hypothesis, which he named
the principle of mechanical transformability, and
appeared in 1918 in his paper on the quantum theory of
line-spectra \cite{bohr1918}. 
Bohr explains that this name indicate in
a more direct way the content of the principle.
This reference on the Bohr paper of 1918 turned the 
Ehrenfest hypothesis widely known but at the same time it
became closely linked to Bohr's work \cite{perez2009}.
The same can be said of the Burger's paper on the
invariance of the phase integrals \cite{perez2009}.

Although the Ehrenfest principle explained the permanence
of a system in a given state, it did not explain why the 
states are discretized. Neither did the Wilson-Sommerfeld
rule as it was introduced as a postulate. The explanation
came with the emergence of quantum mechanics around 1925
which replaced classical mechanics in the explanation
of the motion at the microscopic level. Within quantum
mechanics, the Wilson-Sommerfeld rule was found to be valid
at higher quantum numbers. As to the Ehrenfest principle,
the works of Born \cite{born1927}, Fermi and Persico 
\cite{fermi1926}, and Born and Fock \cite{born1928}
turned it into a theorem of quantum mechanics 
\cite{navarro2006,perez2009}.

\section{Wave mechanics}

The quantization of the electronic orbits of the
hydrogen atom used by Bohr and the quantization
rule used by Sommerfeld explained accurately
the spectrum of the hydrogen including 
the fine structure of the hydrogen lines.
In spite of its successful explanation of 
the spectrum of atoms and several 
problems in atomic physics, the quantum physics up to
1925 was a collection of quantum rules without
a unifying principle \cite{jammer1966}.

In 1925, two quantum theories were proposed which were
latter shown to be equivalent. Heisenberg proposed a
matrix theory \cite{heisenberg1925} and Schrödinger
\cite{schrodinger1926,schrodinger1928} proposed a wave theory. 
The point of departure of the Schrödinger theory
was the relation between the wave theory of light
and geometric optics \cite{jammer1966}. 
Hamilton had shown that there is an analogy between the
principle of least action of mechanics and the
Fermat principle of geometric optics. The principle
of least action is 
\beq
\delta \int 2K dt = 0,
\eeq
where $K$ is the kinetic energy and the minimization 
of the action is subject to trajectories where the
energy $T+V$ is conserved, and can be written in the
form
\beq
\delta \int \sqrt{2m(E-V)}ds = 0.
\label{59}
\eeq
The Fermat principle of geometric optics is
\beq
\delta \int \frac{ds}{v} = 0,
\eeq
where $v$ is the velocity of light. 
Thus the Fermat principle can be regarded as the
principle of least action where $v^{-1}$ plays the
role of the integrand of (\ref{59}) \cite{goldstein1980}. 
As there is a wave theory of light, which reduces
to geometric optics for small wavelength, 
the Schrödinger theory is understood as a
wave theory that reduces to the mechanics.

The wave representation of a quantum theory by 
Schrödinger was suggested by de Broglie
who associated a wave to the motion of a
particle which he called {\it wave phase}
\cite{debroglie1924}. According to de Broglie, 
the wavelength $\lambda$ of the wave associated
to a particle of momentum $p$ is given $p=h/\lambda$,
where $h$ is the Planck constant. The use of wave
naturally leads to quantization.
For instance, the possible states of
a standing wave are the normal modes of vibration,
and the possible values of the wavelengths of a
standing wave forms a discretized set of values.
More generally, the quantization comes from the
fact that the solution of the wave equation 
naturally result in the solution of an
eigenvalue problem, as stated by Schrödinger in
the title of his paper on wave mechanics.

In the first part of his paper on wave mechanics,
Schrödinger introduced the time independent
equation for an electron under the action of the
inverse square force,
\beq
\nabla^2\psi + \frac{2m}{K^2}(E+\frac{e^2}r)\psi = 0,
\eeq
where $K$, according to Schrödinger, must have
the value $K=h/2\pi$ so that the discrete spectrum
corresponds to the Balmer series. 
In the second paper, he considered the one-dimensional 
oscillator whose equation he wrote in the abbreviated
form
\beq
\frac{d^2\psi}{dx^2} + (\frac{a}{\sqrt{b}}-x^2)\psi = 0,
\label{11}
\eeq
and determined the proper values of $a/\sqrt{b}$,
which are $1,3,5,\ldots$ by the use of the
known solution of equation (\ref{11}) in terms of
Hermite orthogonal functions. From this result
the allowed energies of the oscillator are
$E=h\nu(n+1/2)$. 
In the forth part, Schrödinger postulates
that the wave equation is a first order in time
and writes
\beq
\nabla^2\psi - \frac{8\pi^2}{h^2} V\psi \mp \frac{4\pi i}{h}
\frac{\partial\psi}{\partial t} = 0.
\eeq 

In the year following the publication of the
wave theory by Schrödinger there appears 
an approximation method proposed independently 
by Wentzel \cite{wentzel1926}, by Brillouin
\cite{brillouin1926} and by Kramers \cite{kramers1926}.
This approximation corresponds to a perturbation
expansion in powers of the Planck constant.
The zero order approximation gives the classical
result. The first order results in the
Wilson-Sommerfeld rule of the old quantum mechanics. 
The approximation is obtained by
writing the wave function in the form \cite{messiah1966}
\beq
\psi = A e^{i S/\hbar},
\label{90}
\eeq
where $S$ does not depend on $\hbar$
and $A$ is independent of time and is an expansion in powers of $\hbar$.
Replacing it in the time independent Schrödinger equation, 
\beq
-\frac{\hbar^2}{2m} \frac{\partial^2\psi}{\partial x^2} + V \psi = E\psi,
\eeq
the equation containing only terms of order zero
in $\hbar$ is
\beq
\frac{1}{2m} (\frac{\partial S}{\partial x})^2+ V = E,
\label{89a}
\eeq
and the equation coming from terms of first order in
$\hbar$ is
\beq 
\frac{\partial A}{\partial x} \frac{\partial S}{\partial x}
+ \frac{A}{2} \frac{\partial^2 S}{\partial x^2} = 0.
\label{89b}
\eeq

The equation (\ref{89b}) can be integrated with the
result $A^2$ proportional to the reciprocal of 
$(\partial S/\partial x)$.
It is now left to solve the equation (\ref{89a}).
We consider two cases according to the sign of $E-V$. 
If $E\geq V$, we define
\beq
\kappa = \sqrt{2m(E-V)},
\eeq
and the solutions of the equations (\ref{89a})
and (\ref{89b}) are
\beq
S = \pm \int \kappa(x) dx,
\qquad\qquad
A = \frac{1}{\sqrt{\kappa(x)}}.
\eeq
If $E<V$, we define
\beq
\gamma = {2m(V-E)},
\eeq
and the solutions are
\beq
S = \pm i \int \gamma(x) dx,
\qquad\qquad
A = \frac{1}{\sqrt{\gamma(x)}}.
\eeq

Let us suppose that the first condition occurs when
$a\leq x\leq b$, where $V(a)=E$ and $V(b)=E$, are
the classical turning points. The connection of the
solutions at the turning points leads to the
condition \cite{messiah1966}
\beq 
\int_a^b \kappa dx = \hbar\pi(n+\frac12).
\eeq
Considering that the classical momentum $p$ is 
$\kappa$ or $-\kappa$, we may write this condition as
\beq
\oint pdx = h(n+\frac12),
\label{12}
\eeq
which is the Wilson-Sommerfeld rule
except for the 1/2 term.

As the phase integral in the left hand side of equation (\ref{12})
is invariant, it follows that the
quantum number $n$ is an invariant. That is,
if a parameter of the system is slowly varying in 
time, it remains in the same state with the same
quantum number. Nevertheless, a demonstration of
invariance of the quantum state was provided
for the new quantum mechanics, without referring to the 
phase integral. In 1926, one year after the introduction
of the quantum wave by Schrödinger, a demonstration of the
invariance in quantum mechanics was given by Born
\cite{born1927} and by Fermi and Persico \cite{fermi1926}.
Two years later, a more general demonstration was
provided by Born and Fock \cite{born1928} 

\begin{figure}
\centering
\epsfig{file=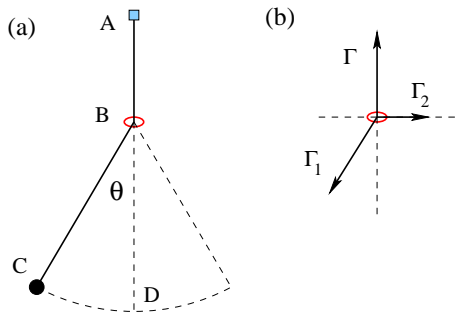,width=6cm}
\caption{(a) The pendulum with variable length.
In the set up employed by Rayleigh the point A is
fixed and the ring B move vertically causing the
variation of the length BC of the pendulum. In
the set up used by Bossut and Lecornu, the ring
B remains immobile and the point A moves vertically.
(b) The forces acting on the ring when it is free
to move upward, where $\Gamma$ is the tension of the
string, $\Gamma_1=\Gamma\cos\theta$, and
$\Gamma_2=\Gamma\sin\theta$.
The net force is upward and equals $T(1-\cos\theta)$.}
\label{pend}
\end{figure}

\section{Rayleigh approach}

\subsection{Pendulum of variable length}

In a paper of 1902 \cite{rayleigh1902}, Rayleigh analyzed
a simple pendulum with its string being varied very slowly. 
Motivated by the theoretical demonstration 
of the radiation pressure by Maxwell and its experimental
confirmation by Lebedev, Rayleigh
inquired whether any other kinds of vibration, such
as sound vibrations, would also cause pressure. 
To answer this question, he posed the problem of finding
the force acted by a vibrating pendulum on its pivot
when its length changes slowly and continuously with time.

The length of the pendulum is changed by the use of
a ring through which passes the string, as shown in
figure \ref{pend}. The ring is constrained to move
vertically and as it moves the length BC of the pendulum
changes although the total length ABC remains constant. 
The problem is to determine the force that tends
to move the ring upwards as the pendulum swings.

The ring is acted by two vertical forces, one of them 
is upward and equal to the tension $\Gamma$ of the string
and the other is downward and equal to $\Gamma\cos\theta$
where $\theta$ is the angle BCD. 
The net upward force on the ring is thus
$F=\Gamma(1-\cos\theta)$. Now the potential energy of the
pendulum is $V=P\ell(1-\cos\theta)$ where $P$ is the
weight of the bob, and $\ell$ is the length BC of the
pendulum. Considering that for small oscillations
$\Gamma$ is approximately equal to $P$, one finds $F=V/\ell$.
As the mean value of the potential energy is one half
of the total energy $E$ of the pendulum, Rayleigh
concludes that the upward mean force $F$ on the ring is
\beq
F = \frac{E}{2\ell}.
\label{6}
\eeq 
As the work done on the ring is equal to decrease in the
energy of the pendulum, then $dE = -F d\ell$, and
$dE = -E d\ell/2\ell$ which by integration gives
$E = a/\sqrt{\ell}$, where $a$ is a constant. 
Although, Rayleigh did not mention it explicitly,
it follows from this result that the quantity 
\beq
I = E\sqrt{\ell}
\eeq
is an invariant quantity
when the length of the pendulum changes slowly with time.
As the frequency of oscillation $\omega$ of a simple 
pendulum executing small oscillations is inversely
proportional to $\sqrt{\ell}$ it follows that
the quantity
\beq
I = \frac{E}{\omega}
\eeq
is an invariant as well.

\begin{figure}
\centering
\epsfig{file=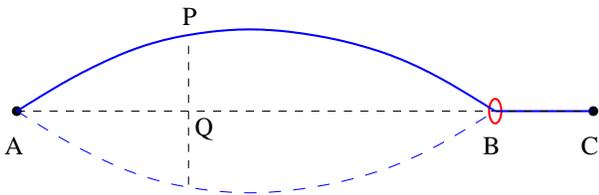,width=8cm}
\caption{The stretched string APBC is fixed at the
points A and C. The ring B restricts the vibration
to occurs only between A and B. The length of the
vibrating string changes by the motion of the ring
along the direction ABC.The string tension remains
unaltered when the position of the ring changes.}
\label{string}
\end{figure}

Rayleigh also treated in the same paper the problem
of the force exerted by a vibrating stretched string
on the points where it is attached.
One end of the stretched string is fixed and 
the other is allowed to move by the use of
a ring as shown in figure \ref{string}.
The position of the ring determines the
length of the vibrating string, which we denote
by $\ell$. Rayleigh argues that the mean force
$F$ acting on the ring is related to the total
energy $E$ of the vibrating string by
\beq
F = \frac{E}{\ell},
\label{7}
\eeq
and is thus equal to the energy per unit
length. Again, the work done on the ring equals
the decrease in energy, 
$dE = -Fd\ell = - Ed\ell/\ell$ and, after
integration, $E = a/\ell$ where $a$ is a
constant, and now $E$ is inversely proportional
to $\ell$. Thus,
\beq
I = E\ell
\eeq
is invariant.

\subsection{Invariance}

The main result of the approach by Rayleigh can be
stated as follows. Let us consider a periodic system
and the force $f$ acted by the system on the environment
at a point which moves as a result of the change of a
parameter $\lambda$. As the system is periodic the
force $f$ oscillates in time but its time average $F$
over one cycle is nonzero. If the parameter changes
slowly $F$ varies slowly and so does the energy $E$
of the system. These two quantities are related by
\beq
\frac{dE}{d\lambda} = - F,
\label{34}
\eeq 
if the parameter changes very slowly with time.
To show this result we proceed as follows.

We consider a system with several degrees of freedom
described by the Lagrangian $L=K-V$, where $K$ is the
kinetic energy and $V$ is the potential energy.
The Lagrangian $L(q,\dot{q},\lambda)$ depends on the
coordinates $q_i$ and velocities $\dot{q}_i$,
which we are denoting collectively by $q$ and $\dot{q}$,
respectively, and on a parameter $\lambda$ which 
depends on time. The Lagrange equations of motion are
\beq
\frac{d}{dt}\frac{\partial L}{\partial\dot{q}_i}
= \frac{\partial L}{\partial q_i}.
\eeq
If we define the momentum
\beq
p_i = \frac{\partial L}{\partial\dot{q}},
\eeq
then the equations of motion are written as
\beq
\frac{dp_i}{dt} = \frac{\partial L}{\partial q_i},
\label{18c}
\eeq
and the differential of $L$ as
\beq
dL = \sum_i p_i d\dot{q}_i + \sum_i \dot{p}_i dq
+ \frac{\partial L}{\partial t}dt.
\eeq
If we perform the Legendre transformation
\beq
H = \sum_i p_i\dot{q}_i - L,
\eeq
then
\beq
dH = \sum_i \dot{q}_i dp_i - \sum_i \dot{p}_i dq
- \frac{\partial L}{\partial t}dt,
\eeq
from which follows the Hamilton equations of motion
\beq
\dot{q}_i = \frac{\partial H}{\partial p_i},
\qquad\qquad
\dot{p}_i = -\frac{\partial H}{\partial q_i},
\label{51}
\eeq
where $H(q,p,\lambda)$ is the Hamiltonian, a function
of the coordinates and momenta, and depends
on time through the parameter $\lambda$.
It also follows that 
\beq
\frac{\partial H}{\partial t} = -\frac{\partial L}{\partial t},
\label{17}
\eeq
where the first derivative is determined at constant
$q$ and $p$ whereas the second at constant $q$ and $\dot{q}$.

The Hamiltonian $H$ equals the total energy $K+V$ if $K$ is
a quadratic form in the velocities. Its variation with time is
\beq
\frac{dH}{dt} = \frac{\partial H}{\partial t},
\label{18e}
\eeq
and as it depends on time through the parameter $\lambda$,
it is not conserved.
Defining the function $f$ by
\beq
f= - \frac{\partial H}{\partial\lambda},
\label{18b}
\eeq
the equation (\ref{18e}) can be written as
\beq
\frac{dH}{dt} = - c f,
\label{18}
\eeq
where $c=d\lambda/dt$ is the rate of variation of the
parameter.

Now we proceed as follows.
We divide the time axis into intervals equal to
the cycle period $T$. The integration of the
equation (\ref{18}) over one cycle beginning
at time $t_0$ and ending at time $t_1$ gives
\beq
\frac{\Delta H}{T} = - \frac1T\int_{t_0}^{t_1} c f(q,p,\lambda)\, dt,
\label{18g}
\eeq
where  
\beq
\Delta H = H(q_1,p_1,\lambda_1) - H(q_0,q_0,\lambda_0),
\eeq
and the subindexes refer to the beginning
and ending of the interval.

We use an approximation
in which the trajectory in phase space is replaced
by a trajectory $(x(t),y(t))$ which is the solution
of the equations of motion by considering that
the parameter $\lambda$ is kept
{\it unchanged} and equal to its value at the beginning
at the interval. The variation of $H$ becomes 
\beq
\Delta H = H(x_1,y_1,\lambda_1) - H(x_0,y_0,\lambda_0).
\eeq
The variation $\Delta\lambda=\lambda_1-\lambda_0$ 
of the parameter in the interval is equal to
$\Delta\lambda=T c$ and $c$
is small because the parameter varies slowly with time.
Therefore $\Delta H/T=c\Delta H/\Delta\lambda$ 
can be approximated by
\beq
c \frac{d}{d\lambda}H(x,y,\lambda),
\eeq
calculated at the beginning at the interval.

Using again the same approximation, we replace
$q$ by $x(t)$, and $p$ by $y(t)$ in the right hand
side of (\ref{18g}). Defining
\beq
F = \frac1T \int_{t_0}^{t_1} f(x,y,\lambda) dt,
\label{50}
\eeq
the equation (\ref{18g}) becomes
\beq
\frac{dE}{d\lambda} = - F,
\label{50b}
\eeq
where $c$ was assumed $c$ to be constant during 
the cycle. In this equation, 
$E$ is equal to $H(x,y,\lambda)$ and depends only 
on $\lambda$ because it is conserved along one cycle,
and the quantity $F$ is understood as the time average
of $f$ along one cycle and also depends only on $\lambda$.
Since $F$ is a function of $\lambda$,
this relation is understood as a differential
equation in $\lambda$. Its solution gives the 
explicit dependence of $E$ on $\lambda$, and
on time since $\lambda$ is a given function of time.

Let us use the notation 
\beq
\overline{f} = \frac1{T}\int_{t_0}^{t_0+T} f dt,
\eeq
for the time average of $f$ over one cycle of period $T$.
The function $f$ depends on the parameter $\lambda$,
which is considered to be fixed. The main result can
then be written in the abbreviated form
\beq
\frac{dE}{d\lambda} = 
\overline{\frac{\partial{\cal H}}{\partial\lambda}},
\label{31}
\eeq
and $E=\overline{\cal H}$. Notice that, as 
${\cal H}$ is conserved if $\lambda$ is fixed,
then $E$ coincides with $E={\cal H}$ during 
one period and this time average is immaterial. 

\subsection{Examples}

Let us apply this approach to the Rayleigh pendulum
The Lagrangian function is given by
\beq
L = \frac12 m\ell^2\dot\theta^2 - \frac12 m g\ell\theta^2.
\eeq
To determined $f=-\partial H/\lambda$, we observe that using the equality
(\ref{17}), it can also be determined by 
$f = \partial L/\partial\lambda$.
Deriving $L$ with respect to $\ell$, we find
\beq
f = m\ell\dot\theta^2 -\frac12 m g\theta^2.
\eeq
The energy $E$ is given by
\beq
E = \frac12 m\ell^2\dot\theta^2 + \frac12 m g\ell\theta^2.
\eeq

The equation of motion, keeping the parameter
$\ell$ unchanged is
\beq
\ell \ddot\theta = -g\theta,
\eeq
whose solution is $\theta=c\cos\omega t$,
where $\omega=\sqrt{g/\ell}$.
Replacing the solution in the expression for $H$
and $f$, and taking the time average over one
cycle, we obtain
\beq
E = \frac12mg c^2 \ell, \qquad\qquad F = \frac14 mg c^2,
\eeq
which gives $F=E/2\ell$, the result (\ref{6}).
The integration of $dE/d\ell=-F$ gives the result 
$E\sqrt{\ell}$ an invariant, or $E/\omega$ an invariant,
results already found.

For a particle of mass $m$ bounded to a spring
of coefficient $k$, the Lagrangian function is
\beq
L = \frac12 m \dot{x}^2 - \frac12 k x^2,
\eeq
and the energy is
\beq
E = \frac12 m \dot{x}^2 + \frac12 k x^2.
\eeq
The equation of motion is $m\ddot{x}=-kx$
whose solution is $x=a\cos\omega t$, where
$\omega=\sqrt{k/m}$. Replacing these results
in the expression for the energy, $E=ka^2/2$.

Let us suppose that the spring coefficient is
the varying parameter. Then $f=-x^2/2$
and $F= -a^2/4$. Therefore, $F=-E/2k$
and the integration of $dE/dk=-F$ gives
$E/\sqrt{k}$ as an invariant, or $E/\omega$.

Suppose now that the mas is the varying parameter.
Then, $f=\dot{x}^2/2$ and $F= k a^2/4m$.
Therefore $F=E/2m$ and the integration of 
$dE/dm=-F$ gives $E\sqrt{m}$ as an invariant,
or $E/\omega$.

Another example consists of a free particle 
of mass $m$ that moves with speed $v$
between two walls that are a distance $\ell$
apart. The mean force $F$ on the wall is the change
of its momentum $2mv$ divided by the time $2\ell/v$ 
between two collisions, that is, $F=mv^2/\ell$
Considering that the kinetic energy is $E=mv^2/2$
one finds $F=2E/\ell$ and the integration
of $dE/dm=-F$ gives $E\ell^2$ as an invariant.

Let us consider now the vibrating stretched
string. Denoting by $u$ the transverse displacement PQ
of the string at the point $Q$ and by $x$ the distance
from this point to the fixed end A, as shown in
figure \ref{string}, the equation of motion for $u$ is
the wave equation
\beq
\frac{\partial^2u}{\partial t^2} 
= c^2 \frac{\partial^2u}{\partial x^2},
\eeq
where $c$ is the velocity of the wave and equal to
$\sqrt{\Gamma/\mu}$, where $\Gamma$ is the tension
of the string and $\mu$ is the mass per unit length.
The wave equation is understood as the equation of
motion associated to the Lagrangian $L=T-V$
where $K$ is the kinetic energy,
\beq
K = \frac{\mu}{2}\int_0^\ell
\left(\frac{\partial u}{\partial t}\right)^2dx,
\eeq
and $V$ is the potential energy,
\beq
V = \frac{\Gamma}2\int_0^\ell
\left(\frac{\partial u}{\partial x}\right)^2 dx.
\eeq
The energy is $H=T+V$.

With the purpose of revealing the explicit
dependence on $\ell$, we change the variable
of integration from $x$ to $y=x/\ell$, with
the result 
\beq
K = \frac{\mu \ell}{2}\int_0^1
\left(\frac{\partial u}{\partial t}\right)^2dy,
\eeq
and $V$ is the potential energy,
\beq
V = \frac{\Gamma}{2\ell}\int_0^1
\left(\frac{\partial u}{\partial y}\right)^2 dy.
\eeq
Using these expression on $L=K-V$, we determine
$f=\partial L/\partial\ell$, which gives the 
following result $f=H/\ell$, where $H=K+V$.
Therefore $F=E/\ell$, which is the result (\ref{7}),
and, using the relation $dE/d\ell=-F$, we find that 
$E=A/\ell$ and $E\ell$ is an invariant as already found.

The solution of the wave equation for
vibrations between the fixed
ends at $x=0$ and $x=\ell$ is the standing wave
\beq
u = A \sin kx\cos\omega t,
\eeq
where $k=n\pi/\ell$ and $n$ is an integer, and
$\omega=kc$. Replacing these results into the
expression for $K$ and $V$ and taking the time
average over one cycle, we find
\beq
E = \mu A^2 c^2 \frac{n^2\pi^2}{4\ell} 
\qquad\qquad  F = \mu A^2 c^2 \frac{n^2\pi^2}{4\ell^2},
\label{35}
\eeq
showing that indeed $E=a/\ell$ and $F=E/\ell$.

\section{Ehrenfest invariant}

\subsection{Invariance}

To demonstrate the invariance of expression (\ref{13})
or its equivalent form (\ref{14}) we proceed as follows.
We start by writing $F$, given by (\ref{50}), in the form
\beq
F = -\frac1T \int_0^T \frac{\partial H}{\partial\lambda}dt,
\eeq
where we have used the expression (\ref{18b}) for $f$,
and we are considering the variables $\lambda$ and
$t$ independent of each other.
Recalling that $E=H$, we obtain
\beq
\frac{dE}{d\lambda} =
\frac1T\!\sum_i\int_0^T \!\!\left(
\frac{\partial H}{\partial p_i}\frac{\partial p_i}{\partial\lambda}
+\sum_i\frac{\partial H}{\partial q_i}\frac{\partial q_i}{\partial\lambda}
+ \frac{\partial H}{\partial\lambda}
\!\right)\!dt.
\eeq
Replacing it in the equation (\ref{50b}), written in the form
\beq
\frac{dE}{d\lambda} + F = 0,
\eeq
we find
\beq
 \sum_i\int_0^T\left(
\frac{\partial H}{\partial p_i}\frac{\partial p_i}{\partial\lambda}
+\frac{\partial H}{\partial q_i}\frac{\partial q_i}{\partial\lambda}
\right) dt = 0.
\eeq
Using the equations of motion (\ref{51}), we find
\beq
 \sum_i\oint\left(\frac{\partial p_i}{\partial\lambda} dq_i
- \frac{\partial q_i}{\partial\lambda} dp_i \right) = 0,
\eeq
which can be written in the form
\beq
\frac{dI}{d\lambda} = 0,
\eeq
where
\beq
I = \frac12\sum_i\oint(p_i dq_i - q_i dp_i ).
\eeq
An integration by parts gives
\beq
I = \sum_i\oint p_i dq_i.
\label{44}
\eeq
Since $dI/d\lambda=0$, it follows that $I$ is indeed an invariant.
We recall that this expression can also be written as
\beq
I = \int_0^T 2K dt,
\label{44b}
\eeq
where $K$ is the kinetic energy and $T$ is the
period of the cycle.

The demonstration just carried out shows that the
invariant (\ref{44}) is a sum terms of the type
\beq
I_i = \oint p_i dq_i.
\label{14c}
\eeq
It does not say whether or not each term $I_i$ is an
invariant. However, if the momentum $p_i$ in this integral
(\ref{14c}) depends only on $q_i$, which is a
statement that the pair of variables $(q_i,p_i)$
is separable from the others, then $I_i$ is an invariant,
which is the result obtained by Burgers. To show this
result, it suffices to write (\ref{14c}) as
\beq
I_i = \int p_i \dot{q}_i dt.
\label{14d}
\eeq
In this form we see that the integrand is twice
the kinetic energy of a system with one
degree of freedom, as no other variables are
involved. But this is the total kinetic energy of
a system with one degree of freedom and is
thus the Ehrenfest invariant.
It is worth mentioning however that the separation of variables
may only occur if an appropriate transformation of variables
is performed.

\subsection{Systems with one and two degrees of freedom}

The Ehrenfest invariant for a system with one degree
of freedom reduces to the phase integral
\beq
I = \oint p dq,
\label{55}
\eeq
where $p$ is the momentum conjugate to $q$.
The general form of a Lagrangian describing a
conservative system with one degree of freedom is
\beq
L = \frac12 m \dot{q}^2 - V,
\eeq
where $m$ might depend on $q$, and $V$ is a function
of $q$ only, and $p=\partial L/\partial\dot{q}$.
As the energy is conserved we
write $E=H(q,p)$ which describes a closed curve
on the phase space. Solving this equation for $p$
and replacing the result in the phase integral, we find
\beq
I = 2 \int \sqrt{2m(E-V)} dq,
\eeq
where the integral is performed in the
interval between the two points of return.

For a particle of mass $m$ under the
action of a harmonic force, the energy is given by
\beq
\frac{p^2}{2m} + \frac12 m\omega^2 x^2 = E,
\eeq
where $\omega$ is the frequency of oscillation.
This equation describes in the phase space an
ellipse of semi-axis equal to $\sqrt{2mE}$
and $\sqrt{2E/m\omega^2}$. The phase integral
is the area of this ellipse and equals
$2\pi E/\omega$, which is thus an invariant.

Another example is given by a free particle
that moves along an axis and collides
with walls that are a distance $\ell$ apart.
The integral (\ref{55}) becomes equal to 
$2p\ell$ which is thus an invariant. 
Taking into account that the kinetic energy
of the particle is $E=p^2/2m$,
it follows that $E \ell^2$ is an invariant
as the wall moves slowly. 

Let us consider now a particle of mass $m$ moving in
a plane under a central force. 
Using polar coordinates, the Lagrangian is given by
\beq
L = \frac12m(\dot{r}^2 + r^2\dot\theta^2) - V,
\eeq
where $V$ is the potential that depends on $r$
but not on $\theta$. The momenta conjugate to
the $r$ and $\theta$ are, respectively,
$p_r=m\dot{r}$ and $p_\theta=mr^2dot\theta$.

One equation of motion is
\beq
\frac{dp_r}{dt} = mr\dot\theta^2 + f,
\eeq
where $f=-dV/dr$ is the centripetal force.
The other equation of motion is $dp_\theta/dt=0$
from which follows that the angular momentum
$p_\theta=mr^2\theta$ is constant. Denoting by
$a$ this constant, then $\dot\theta=a/mr^2$
which replaced in the equation of motion for $r$,
gives
\beq
\frac{dp_r}{dt} = \frac{a^2}{mr^3} + f.
\label{57}
\eeq
This equation tell us that the pair of variables
$(r,p_r)$ is separable and that
\beq
I_1 = \oint p_r dr
\eeq
is an invariant.

To determine $I_1$ explicitly, we multiply (\ref{57})
by $p_r/m=\dot{r}$ and integrate in time to find
\beq
\frac{p_r^2}{2m} + V + \frac{a^2}{2mr^2} = E,
\eeq
where $E$ is a constant. Solving for $p_r$ and replacing
in the integral $I_1$, we get
\beq
I_1 = 2\int \sqrt{2m(E-V)-(a^2/r^2)}\,\, dr.
\eeq

We remark that the integral 
\beq
I_2 = \oint p_\theta d\theta
\eeq
is also an invariant, in fact it is
a constant, $I_2=2\pi a$.

\subsection{Particle under a central force}

We wish to determine now the invariants of a
system with three degrees of
freedom corresponding to a particle under the
action of a central force as is the case of the
Kepler problem where this force is proportional
to inverse of the square of the distance
\cite{born1927b,landau1960,terhaar1964,borghi2013}.
In spherical coordinates, the
variables are separable and for each pair
of conjugate variables there corresponds
an invariant of the form (\ref{55}).
Using spherical coordinates $r$, $\theta$,
and $\phi$, with origin at the center of
force, the kinetic energy is given by
\beq
K = \frac{m}2\dot{r}^2 + \frac{m}2 r^2\dot\theta^2
+ \frac{m}2 r^2 \dot\phi^2\sin^2\theta,
\eeq
and the Lagrangian is $L=T-V$ where
$V(r)$ is the potential energy that depends on $r$ only.

The conjugate momenta are
\beq
p_0 = m \dot{r},
\eeq
\beq
p_1 = m r^2\dot\theta,
\eeq
\beq
p_2 = m r^2\sin^2\theta\dot\phi,
\eeq
and the equations of motion are
\beq
\frac{dp_0}{dt} = m r\dot\theta^2 + mr\sin^2\theta\dot\phi^2 + f,
\label{52}
\eeq
\beq
\frac{dp_1}{dt} = m r^2\dot\phi^2\sin\theta\cos\theta,
\eeq
\beq
\frac{dp_2}{dt} = 0,
\eeq
where $f=-dV/dr$ is the central force that depends on $r$ only.
From the last equation it follows that $p_2$ is constant.
Setting this constant equal to $a$, then $\dot\phi=a/mr^2\sin^2\theta$,
which replaced in the equation of motion for $p_1$ gives
\beq
\frac{dp_1}{dt} = \frac{a^2\cos\theta}{mr^2\sin^3\theta}.
\eeq
Multiplying this equation by $p_1=mr^2\dot\theta$
and integrating in time, we find
\beq
p_1^2 + \frac{a^2}{\sin^2\theta} = b^2,
\label{52a}
\eeq
where $b$ is a constant. Therefore $p_1$ in the
integral
\beq
I_1 = \oint p_1 d\theta
\eeq
depends only on $\theta$ and is an invariant.

Let us write equation (\ref{52}) in the form
\beq
\frac{dp_0}{dt} = \frac{p_1^2}{mr^3} + \frac{p_2^2}{mr^3\sin^2\theta} + f.
\eeq
Replacing the result (\ref{52a}) in this equation and bearing
in mind that $p_2=a$, we find
\beq
\frac{dp_0}{dt} = \frac{b^2}{mr^3} + f.
\label{58}
\eeq
Multiplying by $p_0/m=\dot{r}$ and integrating in time,
we find
\beq
\frac{p_0^2}{2m} + V + \frac{b^2}{2mr^2} = E,
\label{58a}
\eeq
where $E$ is a constant. Since $p_0$ depends only
on $r$, it follows that
\beq
I_0 = \oint p_0 dr
\eeq
is an invariant.
Solving equation (\ref{58a}) for $p_0$ and replacing in
this integral we reach the result
\beq
I_0 = 2\int \sqrt{2m(E-V)-(b^2/r^2)}\,\, dr.
\eeq

We remark that the motion of a particle under a central
force is restricted to the plane defined by the velocity
and the center of force. Therefore the problem could be
reduced two a system with two degrees of freedom like we
have done previously. However, there might be parameters
that could remove the motion from this plane. In this
case it is necessary to consider the problem in three
dimensions as we have just done.

\subsection{Particle on a magnetic field}
 
Let us consider the motion of a particle of mass $m$
and charge $e$ in a uniform magnetic field $B$ which
is parallel to the $z$ axis. The Lagrangian $L$ in cylindrical
coordinates is given by \cite{terhaar1966}
\beq
L = \frac12 m(\dot{r}^2 + r^2\dot{\theta}^2 + \dot{z}^2)
+ \frac12 e B r^2 \dot{\theta}.
\eeq
The momenta conjugate to $r$, $\theta$ and $z$ are,
respectively,
\beq
p_0=m\dot{r},
\eeq
\beq
p_1=mr^2\dot\theta + \frac12 e B r^2,
\eeq
\beq
p_2=m\dot{z},
\eeq
and the equations of motion are
\beq
\frac{dp_0}{dt} = m r \dot\theta^2 + eB r \dot\theta,
\eeq
$dp_1/dt = 0$, and $dp_2/dt=0$. From these two last
equations, it follows that $\dot{z}=b$, a constant,
and that
\beq
mr^2\dot\theta + \frac12 e B r^2 = a,
\eeq
where $a$ is another constant, or
\beq
\dot\theta  = \frac{a}{mr^2} - \frac{e B}{2m}.
\eeq
Replacing this last result in the equation of motion
for $p_0$ we find
\beq
\frac{dp_0}{dt} = \frac{a^2}{mr^3} - \frac{e^2 B^2 r}{4m}.
\eeq
Multiplying this equation by $p_0/m=\dot{r}$
and integrating in time, we find
\beq
\frac{p_0^2}{2m} + \frac{a^2}{2mr^2} + \frac{e^2 B^2 r^2}{8m} = E,
\eeq
where $E$ is a constant. As $p_0$ depends only on $r$,
it follows that the phase integral
\beq
I_0 = \oint p_0 dr
\eeq
is an invariant.

\section{Dynamic approach}

The approaches to the mechanical
problem of parametric action treated 
up to now involve an
approximation in which the parameter is
held constant while the system completes
a full cycle. This is the case of the
Rayleigh approach just presented as well
as that of Ehrenfest. We may say that the
parameter varies in time in steps, the
time of each step being equal to the
period of a cycle in which the parameter
is held constant. In other word, the
parameter as a function of time looks
like a staircase. Nevertheless, these approaches give
correct results in the limit of infinitely
slow variation of the parameter. 

In the following, the problem is treat
without considering the parameter fixed
in a cycle but still considering that
the variation of the parameter is slow.
In other word, the parameter varies
continuously in time rather than increasing
by steps as was the case of the Rayleigh and 
of the Ehrenfest approaches.

We wish to determine the properties of a
system in the regime of slow parametric action
which is defined as follows. Let a parameter 
$\lambda$ varies linearly in time, that is,
$\lambda=\lambda_0(1+\varepsilon t)$
where $\varepsilon$ is small. This regime
is defined for times smaller that $1/\varepsilon$
and a quantity is an invariant if it varies
little in this interval \cite{arnold1963,arnold1978}.

\subsection{Pendulum of variable length}

The problem of the pendulum with variable length 
was treated by Lecornu in 1895 \cite{lecornu1895}
and previously by Bossut in 1778 \cite{bossut1778}, 
although they did not draw the relevant conclusion 
of Rayleigh concerning the relation between
energy and the length of the pendulum. 
Bossut imagined the oscillations
of an unguided bucket during its ascent in a mine well.
Bearing mind the figure \ref{pend}, the problem 
is formulated by considering that the ring is fixed and that
the string is moved upward.
Using this set up, Bossut and Lecornu otained
the equation of motion.
According to Lecornu, Bossut reduced the differential
equation of the second order into a Riccati equation but 
he did not give a solution. Lecornu reduced the
the differential equation of motion to an equation that
could be solved through the use of Bessel functions. 
The problem was examined later on in 1923 by Krutkov and
Fock \cite{krutkow1923} who demonstrated the invariance of $E/\omega$
by using the asymptotic form of the Bessel function.
In the following we present the treatment of 
the problem following the treatment of Lecornu and
Krutkow and Fock.
More recently, the problem has been treated by
Sánchez-Soto and Zoido \cite{sanchez2013}.

Let $x$ and $y$ be the projections of CB in figure \ref{pend}
along the horizontal and vertical directions, respectively.
They are related to length $\ell$ of the pendulum
and the angle $\theta$ by
\beq
x= \ell\sin\theta,
\label{7a}
\eeq
\beq
y= \ell\cos\theta,
\label{7b}
\eeq
and $\ell$ is a given function of time $t$.
We suppose that the length of the pendulum
to vary linearly with time, $\ell=\ell_0+ht$.
The kinetic energy of the pendulum is
\beq
K = \frac12m(\dot{x}^2 + \dot{y}^2),
\eeq
where $m$ is the mass of the bob.
From the expression of $x$ and $y$ we find
\beq
\dot{x} = \ell\dot{\theta}\cos\theta + h \sin\theta,
\eeq
\beq
\dot{y} = -\ell\dot{\theta}\sin\theta + h \cos\theta,
\eeq
which replaced in the expression for $K$ gives
\beq
K = \frac12 m\ell^2\dot{\theta}^2 + \frac12 m h^2.
\label{12a}
\eeq
Notice that the second term is the kinetic energy
due to the steady vertical motion of the bob with
velocity $h$. 
The potential energy is $V = mg(\ell_0-\ell\cos\theta)$, 
where $g$ is the acceleration of gravity.  
As we will consider only small oscillations, $\theta$ is
small and we may write
\beq
V = \frac12mg\ell\,\theta^2 + mg(\ell_0-\ell).
\label{12b}
\eeq
Notice that the second term is the potential energy
related to the vertical motion of the bob.

The equation of motion is derived from the Lagrangian equation
\beq
\frac{d}{dL}\frac{\partial L}{\partial\dot{\theta}} =
\frac{\partial L}{\partial\theta},
\eeq 
where $L=K-V$ is the Lagrangian, given by
\beq
L = \frac12 m \ell^2\dot{\theta}^2 - \frac12mg\ell\,\theta^2
+ \frac12 m h^2 - mg(\ell_0-\ell).
\eeq
From this expression we reach the equation of motion
\beq
\ell\ddot{\theta} + 2h\dot{\theta} =  - g\theta,
\label{8}
\eeq
valid for small oscillations.
Changing variable from $t$ to $\ell=\ell_0+ht$,
this equation becomes
\beq
\ell\frac{d^2\theta}{d\ell^2} + 2\frac{d\theta}{d\ell}
+ \frac{g}{h^2}\theta = 0,
\label{10a}
\eeq
which is the equation derived by Lecornu \cite{lecornu1895}.

It is convenient to define the variable $s=\ell/g$
or $s=at+b$, where $a=h/g$ and $b=\ell_0/g$, 
from which we may write the equation of motion as
\beq
s\frac{d^2\theta}{ds^2} + 2\frac{d\theta}{ds}
+ \frac\theta{a^2} = 0.
\label{10}
\eeq
Performing the change of variables defined by 
$z=2\sqrt{s}/a$ and $\phi=z\theta$ we reach the
equation
\beq
z^2\frac{d^2\phi}{dz^2} + z\frac{d\phi}{dz} + (z^2-1)\phi = 0.
\eeq
In this form, we see that the solutions are the Bessel
functions of first order $J_1(z)$ and $Y_1(z)$
\cite{abramowitz1965}, that is,
\beq
\phi= A_1 J_1(z) + A_2 Y_1(z),
\eeq
where $A_1$ and $A_2$ are constant.
\beq
\theta = \frac1z[A_1 J_1(z) + A_2 Y_1(z)],
\eeq
which gives $\theta$ as a function of $t$
if we recall that $z=2\sqrt{s}/a$ and that $s=at+b$.

As we wish to get the solution for a very slow variation of
the length of the pendulum, which means that $a$ is very small,
it suffices to consider the solution for large values of $z$.
For as $z=2\sqrt{s}/a$ and considering a 
finite value of $s=at+b$, $z$ will increase as $1/a$.
Therefore, we use the asymptotic expression of the Bessel
functions \cite{abramowitz1965},
as did Trutkov and Fock \cite{krutkow1923}, namely
\beq
J_1(z) = \left(\frac{2}{\pi z}\right)^{1/2}
\sin\left(z-\frac\pi4\right),
\eeq
\beq
Y_1(z) = \left(\frac{2}{\pi z}\right)^{1/2}
\cos\left(z-\frac\pi4\right).
\eeq
The solution can thus be written as
\beq
\theta = cs^{-3/4} \cos\left(\frac2a\sqrt{s}-\frac{\pi}4\right).
\label{15}
\eeq

The energy $E$ of the pendulum is the first part
of the kinetic energy given by (\ref{12a}) plus the
first part of the potential energy given by (\ref{12b}),
\beq
E = \frac{m}2 \left(\ell^2\dot{\theta}^2 + \ell g\theta^2\right),
\eeq
which can be written as
\beq
E = \frac{m}2 g^2\left[
a^2 s^2 \left(\frac{d\theta}{ds}\right)^2 + s\theta^2\right].
\label{19}
\eeq
Replacing the solution (\ref{15}) in this equation,
 we reach the following expression for the energy
\beq
E = \frac{mg^2c^2}{2\sqrt{s}}
= \frac{mgc^2}{2} \sqrt{\frac{g}{\ell}},
\label{19a}
\eeq
where we have neglected terms of order equal or larger
that $a$.
That is, the energy of the pendulum is proportional
to the inverse of $\sqrt{\ell}$, the Rayleigh relation.
Bearing in mind that the frequency is $\omega=\sqrt{g/\ell}$,
we may wright
\beq
E = \frac{mgc^2}{2} \omega,
\label{19b}
\eeq
and $E/\omega$ is an adiabatic invariant.

In the treatment that we have just given to the 
pendulum with variable length, we have
used the set up employed by Bossut and Lecornu,
which corresponds to keep the ring of figure
\ref{pend} fixed, while the point A moves
vertically. In the original set up
of Rayleigh, the point A is kept fixed while
the ring B moves vertically. In this case
the origin of the axis $y$ should
be placed at the point A, which is fixed,
rather than at the point B, which moves. 
The relation between $y$ and the angle $\theta$
becomes
\beq
y = \ell_0 - \ell + \ell\cos\theta.
\eeq
The axis $x$ remains the same and given by
(\ref{7b}).

It is straightforward to show that, for small
oscillations and for $\ell=\ell_0+ht$, the equation
of motion for $\theta$ for the Rayleigh set up is
identical to the Bossut set up, given by equation
(\ref{8}). The kinetic and potential energies for
the Rayleigh set up are given by the first parts
of equation (\ref{12a}) and (\ref{12b}), respectively,
since for the Rayleigh set up there is no vertical
net translation, the total energy being given by
(\ref{19}), with the results (\ref{19a}) and (\ref{19b}).

\subsection{Harmonic oscillator of variable frequency}

Let us consider a harmonic oscillator 
along the $u$ axis with 
variable mass $m$ and variable spring coefficient $k$.
The equation of motion is
\beq
\frac{d}{dt}(m\dot{u})= - k u.
\eeq
If the mass varies linearly in time, $m=m_0+\mu t$,
and $k$ is constant the equation of motion
reduces to
\beq
m\ddot{u} + \mu\dot{u} = - k u.
\eeq
This equation is identical to the equation
(\ref{8}) and can thus be solved in like manner.

We consider now that $m$ is
constant and that the spring coefficient $k$ varies
varies with time \cite{kulsrud1957,wells2006,robnik2006}.
In this case the equation of motion reduces to
\beq
\ddot{u}= - s u,
\label{21}
\eeq
where $s=k/m$ From now on we suppose $k$ varies linearly
with time, that is, $s= b + a t$ where $a$ is a small
quantity. Changing variable from $t$ to $s$ we get
\beq
a^2\frac{d^2u}{ds^2}= - s u.
\eeq
Making another change of variable
from $s$ to $z=a^{-2/3}s$, we reach 
the following equation
\beq
\frac{d^2u}{dz^2}= - z u.
\eeq

The solutions of this equation are the Airy
functions ${\cal A}i(-z)$ and ${\cal B}i(-z)$ 
\cite{abramowitz1965},
\beq
u = c_1 \,{\cal A}i(-z) + c_2 \,{\cal B}i(-z),
\eeq
where $c_1$ and $c_2$ are constants.
As we wish to get the solution for a very slow
variation of the spring coefficient, which 
means that $a$ is small, and bearing in
mind that $z=s/a^{2/3}$,
it suffices to
consider the solutions for large values of $z$.
The asymptotic forms of the Airy functions are
\cite{abramowitz1965}
\beq
{\cal A}i(z) = \frac1{\sqrt{\pi}} z^{-1/4}
\cos(\frac23 z^{3/2}-\frac\pi4),
\eeq
\beq
{\cal B}i(z) = -\frac1{\sqrt{\pi}} z^{-1/4}
\sin(\frac23 z^{3/2}-\frac\pi4).
\eeq
The solution can thus be written as
\beq
u = c s^{-1/4} \cos\left(\frac{2}{3a} s^{3/2} +c_0\right),
\label{20}
\eeq
where $c$ and $c_0$ are constants and  we recall that
$s$ is a function of time, $s=b+a t$.

The energy $E$ of the harmonic oscillator is
\beq
E = \frac{m}2 \dot{u}^2 + \frac{k}2 u^2.
\label{30}
\eeq
From this expression and using the asymptotic solution,
we find
\beq
E = \frac{mc^2}2 \sqrt{s},
\eeq
where we have neglected terms of the order equal or
greater than $a/s$.
Recalling that $\sqrt{s}$ can be understood as the
frequency $\omega=\sqrt{k/m}=\sqrt{s}$, it follows
that the energy of a harmonic oscillator of 
variable frequency is proportional to the 
frequency, or that the ratio $E/\omega$ is 
an invariant.

\subsection{General time dependence}

We ask whether an expression of the type 
\beq
u = r(t) \cos \theta(t),
\label{29}
\eeq
could be the solution of the equation of motion
for the harmonic oscillator of variable spring
coefficient. If we replace the expression
(\ref{29}) in the equation (\ref{21}) we 
find that it is indeed a solution as long
as the following equations involving
$r$, $\theta$ and $s$ are satisfied
\cite{kulsrud1957}:
\beq
2 \dot{r}\omega + r \dot{\omega} = 0,
\eeq
\beq
\ddot{r} - r \omega^2 + s r = 0,
\eeq
where $\dot{\theta}=\omega$.
The solution of the first equation gives
\beq
r=c\,\omega^{-1/2},
\eeq
where $c$ is an arbitrary constant and 
\beq
s = \omega^2 - \frac{\ddot{r}}{r}.
\label{33}
\eeq
Therefore, given $\omega$ as a function of time,
we determine $r$ and then $s$. By the integration
of $\dot{\theta}=\omega$, we determine $\theta$.

Replacing the solution (\ref{29}) into the
expression (\ref{30}), and bearing in mind that
$k/m=s$, we find 
\beq
E = \frac{mc^2}2\omega,
\eeq
where we have neglected terms of the order equal or
larger than $\dot{\omega}/\omega$. Again we find that
$E/\omega$ is an invariant.

A simplification arises if we suppose that $\omega$
is a finite function of $at+b$ where $a$ is a small
quantity. In this case $\ddot{r}/r$ will be of the
order $a^2$ and can be neglected in the expression
(\ref{33}), which reduces to
\beq
s=\omega^2.
\label{33a}
\eeq 
A possible solutions for the dependence of
$\omega$ with time is
\beq
\omega=at+b,
\eeq
which gives 
\beq
\theta=\frac12 at^2+bt + c_0,
\qquad
r = c (at+b)^{-1/2},
\eeq
and, using (\ref{33a}),
\beq
s = (at+b)^2.
\eeq
Another solution is 
\beq
\omega=(at+b)^{1/2},
\eeq
which gives
\beq
\theta =\frac2{3a}(at+b)^{3/2} + c_0,
\qquad
r = c (at+b)^{-1/4},
\eeq
and, using (\ref{33a}),
\beq
s = at+b.
\eeq
This solution is identified with that given by
equation (\ref{20}) if we recall that in (\ref{20}),
$s$ equals $at+b$.
Yet another solution is
\beq
\omega = (at+b)^{-1/2},
\eeq
which gives
\beq
\theta = \frac2a (at+b)^{1/2} + c_0,
\qquad
r = c(at+b)^{1/4},
\eeq
and, using (\ref{33a}),
\beq
s = \frac1{at+b}.
\eeq
This solution is identified with that given by
equation (\ref{15}) if we recall that in equation
(\ref{15}) $s$ equals $at+b$.

\subsection{Vibrating string of variable length}

Let us denote by $\phi$ the transverse displacement PQ
of the string at the point Q and by $x$ the distance
from this point to the fixed end A, as shown if 
figure \ref{string}. The equation
of motion for a uniform string is the wave equation
\beq
\frac{\partial^2\phi}{\partial t^2}
= c^2 \frac{\partial^2\phi}{\partial x^2},
\eeq
where $c$ is the velocity of the wave and equal
to $\sqrt{\Gamma/\mu}$ where $\Gamma$ is the tension of
the string and $\mu$ is the mass per unit length.
The vibration
occurs only for $0\leq x\leq \ell$ where $\ell$ is
the distance of the ring B to the fixed end A.
The boundary conditions are $\phi=0$ for $x=0$ and
$x=\ell$. We wish to solve this equation as
the length changes with time and determine
the energy $E$ of the vibrating string which is given by
\beq
E = \frac\mu2 \int_0^\ell \left(\frac{\partial\phi}{\partial t}\right)^2 dx
+ \frac{\Gamma}2 \int_0^\ell \left(\frac{\partial\phi}{\partial x}\right)^2 dx.
\label{35a}
\eeq
A closed solution of the wave equation can be obtained
for the following time dependence of the string length
$\ell = \ell_0/(1+\varepsilon t)$.

To solve the wave equation as $\ell$
varies with time we proceed as follows. 
We assume a solution of the type
\beq
\phi = e^{i(ax^2+\theta)} \sin k x,
\eeq
where $a$ is a constant and $\theta$ is a function
of $t$. The coefficient $k$ is chosen to be equal to
$n\pi/\ell$ where $n$ is an integer so that $\phi$
vanishes at $x=0$ and $x=\ell$, as desired.
As $\ell$ depends on time, so does $k$, that is,
\beq
k = \frac{n\pi}{\ell}= \frac{n\pi}{\ell_0}(1+\epsilon t).
\eeq
Replacing the
solution into the wave equation and bearing in mind that
$k$ depends on time, we find 
$a=k_0\varepsilon/2c$, where $k_0=n\pi/\ell_0$,
and $\dot{\theta} = k c$, which by integration gives
\beq
\theta = k_0 c(t+\frac12\varepsilon t^2).
\eeq

The other solution corresponds to the complex
conjugate of this expression. Since $\phi$
is real, we sum the two solutions to obtain
\beq
\phi = A \cos \left(\frac{k_0\varepsilon x^2}{2c}
+ k_0ct+\frac{k_0c \varepsilon t^2}2 \right)\sin kx.
\eeq
Replacing this result in the expression (\ref{35a})
for the energy $E$, we obtain the result (\ref{35})
found before.

\section{Quantum mechanics}

\subsection{Parametric invariance}

Born and Fock \cite{born1928}, in their paper
of 1928 stated the invariance in the following terms. 
If the system was in a certain
state described by a certain quantum number,
the probability to change the state, by a slow variation of
a parameter, is infinitely small, in spite of the
fact that the change in the energy levels be of
finite amount. They considered
a discrete and a non-degenerate spectrum of energies
except for the accidental degeneracy due to crossing
of two energy eigenvalues. Demonstration of
the invariance with less restrictions was given 
by Kato in 1950 \cite{kato1950}. Other demonstrations
and discussions of invariance in quantum mechanics
are found in several papers
\cite{hwang1977,nenciu1980,narnhofer1982,avron1999,wu2005,bachmann2017}
and books on quantum mechanics
\cite{messiah1966,gasiorowicz1974,griffiths1994,sakurai2010}.

In the following, we show that if the variation of
a parameter is infinitely slow the system remains
in the same quantum state. To this end we use an
approach analogous to the one we employed above 
for the classical case. We start by consider a system
described by the Schrödinger equation
\beq
i\hbar \frac{\partial\psi}{\partial t} = {\cal H}_\lambda\psi,
\label{60}
\eeq
where $\psi$ is the wave function and ${\cal H}_\lambda$ is the
Hamiltonian operator, which depends on a parameter
$\lambda$ which depends on time.

Let us consider the following quantity 
\beq
H = \la \psi|{\cal H}_\lambda |\psi\ra,
\eeq 
where $\psi$ is a solution of the 
equation (\ref{60}).

As the Hamiltonian depends explicitly on time
through the parameter, $H$ is not conserved.
Its variation in time is
\beq
\frac{dH}{dt}
= \la \psi| \frac{\partial{\cal H}_\lambda}{\partial t}|\psi\ra,
\eeq
which we write in the form
\beq
\frac{dH}{dt} = - c \la \psi|{\cal F}_\lambda|\psi\ra,
\label{74}
\eeq
where $\cal F$ is the operator
\beq
{\cal F}_\lambda = - \frac{\partial{\cal H}_\lambda}{\partial\lambda},
\eeq
and $c=d\lambda/dt$.

Now let $\phi_\lambda$ be a solution of the
Schrödinger equation (\ref{60}) with the condition that the
parameter $\lambda$ is kept {\it unchanged}. We define the
following quantities
\beq
E = \la \phi_\lambda|{\cal H}_\lambda|\phi_\lambda\ra,
\eeq
and
\beq
F = \frac1T \int_0^T \la \phi_\lambda|{\cal F}_\lambda|\phi_\lambda\ra.
\eeq
In accordance with the reasoning given above,
for the classical case, the approximation amounts
to replace $\psi$ by $\phi_\lambda$. The resulting
equation is
\beq
\frac{dE}{d\lambda} = - F.
\label{72}
\eeq

We now write this equation in the form
\beq
\int_0^T  \left(\frac{d}{dt}\la \phi_\lambda|{\cal H}_\lambda|\phi_\lambda\ra
-\la \phi_\lambda|{\cal F}_\lambda|\phi_\lambda\ra\right) dt = 0.
\label{71}
\eeq
The equation (\ref{71}) can be written in the following
equivalent form
\beq
\int_0^T\left(
\la \frac{\partial \phi_\lambda}
{\partial\lambda}| {\cal H}_\lambda |\phi_\lambda\ra 
+ \la \phi_\lambda| {\cal H}_\lambda | \frac{\partial \phi_\lambda}
{\partial\lambda}\ra \right)dt = 0.
\label{75}
\eeq

Let us now denote by $\phi_{\lambda n}$ and $E_{\lambda n}$
the eigenfunctions and eigenvalues of of ${\cal H}_\lambda$,
where $\lambda$ is considered to be fixed. We may then
expand $\phi_\lambda$,
\beq
\phi_\lambda = \sum_n a_{\lambda n} \phi_{\lambda n},
\eeq
We may also expand the derivative of $\phi_\lambda$
with respect to $\lambda$,
\beq
\frac{\partial \phi_\lambda}{\partial\lambda} =
\sum_n b_{\lambda n} \phi_{\lambda n}.
\eeq
Replacing these expansions in equation (\ref{75}), we find
\beq
\sum_n \int_0^T E_{\lambda n} (a_{\lambda n} b_{\lambda n}^*
+ a_{\lambda n}^* b_{\lambda n})dt = 0,
\eeq
where we have assumed that the eigenfunctions are
orthonormalized. A solution of this equation corresponds to
the case where the coefficients are all zero except one
of them, in which case the expression between parentheses
vanishes. Therefore if the system is initially in a certain
state, say state $\phi_{\lambda n}$ it remains in this
state as $\lambda$ is varied slowly. In other words,
the quantum number $n$ is invariant.

\subsection{Electron on a rotating field}

Let us consider the evolution of the spin
of an electron in a rotating magnetic field
\cite{griffiths1994,wu2005}.
The $x$ and $y$ components of the magnetic field
are $B\cos\theta$ and $B\sin\theta$
where $\theta$ is the time dependent parameter,
$\theta=\omega t$.
In the representation where the component $z$ of
the electron spin is diagonal, the Hamiltonian
is given by the square matrix
\beq
{\cal H} = \mu B (\sigma_x \cos\theta + \sigma_y \sin\theta),
\eeq
where $\mu=e\hbar/2m$ is the Bohr magneton and
$\sigma_x$ and $\sigma_y$ are the Pauli matrix.

We define by $\chi_+$
and $\chi_-$ the basis where the Pauli matrix
$\sigma_z$ is diagonal, which are
the column matrices with elements
1 and 0, and 0 and 1, respectively.
The eigenvectors and eigenvalues of ${\cal H}$ are 
\beq
\phi_1 = \frac1{\sqrt{2}}(e^{-i\theta/2}\chi_+ 
+ e^{i\theta/2}\chi_-), \qquad \varepsilon_1 = \mu B,
\eeq
\beq
\phi_2 = \frac1{\sqrt{2}}(e^{-i\theta/2}\chi_+ 
- e^{i\theta/2}\chi_-), \qquad \varepsilon_2 = - \mu B,
\eeq
It is useful to know that
\beq
\frac{d\phi_1}{dt} = - \frac{i\omega}2 \phi_2,
\eeq
\beq
\frac{d\phi_2}{dt} = - \frac{i\omega}2 \phi_1.
\eeq

The Schrödinger equation is
\beq
i\hbar \frac{d\chi}{dt} = {\cal H} \chi,
\eeq
where $\chi$ is the spinor. 
Writing $\chi = x \phi_1 + y \phi_2$, 
the Schrödinger equation becomes
\beq
i \frac{dx}{dt} + \frac{\omega}2 y = \frac{\omega_c}2 x,
\eeq
\beq
i \frac{dy}{dt} + \frac{\omega}2 x = -\frac{\omega_c}2 y,
\eeq
where $\omega_c=2\mu B/\hbar= e B/m$ is the cyclotron frequency.
The solution for the case $x=1$ and $y=0$ for
$t=0$ is
\beq
x = \cos\gamma t - \frac{i\omega_c}{2\gamma} \sin \gamma t,
\eeq
\beq
y = \frac{i\omega}{2\gamma}\sin\gamma t,
\eeq
where
\beq
\gamma = \frac12\sqrt{\omega^2+\omega_c^2}.
\eeq

We remark that $|x|^2+|y|^2=1$ so that $\chi$
is normalized. The probability of the system
to be found in the state $\phi_1$ and $\phi_2$
are respectively
\beq
|x|^2 = \cos^2\gamma t + \frac{\omega_c^2}{4\gamma^2}\sin^2\gamma t,
\eeq
\beq
|y|^2 = \frac{\omega^2}{4\gamma^2}\sin^2\gamma t.
\eeq
In the regime $\omega/\omega_c\ll 1$, the
probability to change from state 1 to state 2
is very small, of the order of $(\omega/\omega_c)^2$,
and we recall that $\omega$ is the rate of change
of the parameter $\theta=\omega t$.

\subsection{Square well with a moving wall}

We consider a particle confined in a one-dimensional
box which is equivalent to the
motion of a particle under an infinite square well
potential. One wall of the box is
fixed and the other moves linearly. Denoting by
$\ell$ the length of the box, we assume that  
$\ell=\ell_0(1+\varepsilon t)$. A closed solution of the 
Schrödinger equation can be found for this
case and in fact
a solution was given by Doescher and Rice \cite{doescher1969}.
In the following we present the solution for
this problem.

The Schrödinger equation to be solved is
\beq
i\hbar \frac{\partial \psi}{\partial t} = - \frac{\hbar^2}{2m}
\frac{\partial^2\psi}{\partial x^2},
\eeq
with the boundary condition that $\psi$ vanishes at $x=0$ and
at $x=\ell$. 

We assume a solution of the type
\beq
\psi =
\sqrt{\frac2\ell}e^{i(\alpha x^2+\theta)}\sin kx,
\label{70}
\eeq
where $\alpha$ and $\theta$ are functions
of $t$, and $k=n\pi/\ell$ so that $\psi$ vanishes at $x=0$
and $x=\ell$, as required.
Replacing this expression in the Schrödinger equation, we find
\beq
\alpha = \frac{m}{2\hbar}\frac{\varepsilon}{1+\varepsilon t},
\eeq
\beq
\dot\theta = - \frac{\hbar}{2m}k^2,
\eeq
which integrated gives
\beq
\theta = - \frac{\hbar k_0^2 t}{2m(1+\varepsilon t)},
\eeq
where $k_0=n\pi/\ell_0$.

Replacing the above results in the equation (\ref{70}),
we reach the following expression \cite{doescher1969}  
\beq
\psi = \sqrt{\frac2\ell}\exp\{
\frac{i}{\hbar} \frac{m^2 \varepsilon x^2
- \hbar^2 k_0^2 t}{2 m (1+\varepsilon t)}
\}\sin kx.
\eeq

\subsection{Hamiltonian obeying a scaling relation}

A closed solution can also be provided when the Hamiltonian
${\cal H}_\lambda (x)$ that depends on a parameter
$\lambda$ obeys the scaling relation 
\beq 
{\cal H}_\lambda (x) = \lambda^b \hat{\cal H}(\xi)
\qquad\qquad \xi = \lambda^a x,
\label{87a}
\eeq 
where $\hat{\cal H}(\xi)$ only on $\xi$ but
not on $\lambda$.
The eigenfunctions $\hat\phi(\xi)$ and the eigenvalues $\hat{E}$
of $\hat{\cal H}(\xi)$ are related to the eigenfunctions
$\phi_\lambda (x)$ and eigenvalues $E$ of the original Hamiltonian by
\beq
E_\lambda = \lambda^b \hat{E},
\label{87b}
\eeq
\beq
\phi_\lambda (x) = \lambda^{a/2} \hat\phi(\xi).
\label{87c}
\eeq
This last relation follows from the normalization of
the eigenfunctions.

A scaling relation of this type is obeyed by the 
Hamiltonian describing a particle in a box, 
in which case $a=-1$ and $b=-2$, and $\lambda$ is
the length of the box. It is also obeyed by the
Hamiltonian of the harmonic oscillator
in which case $a=1/2$ and $b=1$, and $\lambda$ is
the frequency of the oscillation.

We consider the Schrödinger equation
\beq
i\hbar \frac{\partial\psi}{\partial t} = {\cal H} \psi,
\label{60a}
\eeq
where we are omitting the index $\lambda$ in the
Hamiltonian ${\cal H}(x)$, which depends on time
through the parameter $\lambda$.
The Hamiltonian operator
\beq
{\cal H} = {\cal K} + {\cal V}
\eeq
is the sum of the kinetic energy 
operator ${\cal K}$ and ${\cal V}$ is 
the potential energy. In the 
position representation, which we use here,
${\cal V}$ is a multiplying operator, that is,
a function of $x$.

Let $\phi(x)$ be one of the eigenfunctions
of ${\cal H}$ and $E$ the corresponding
eigenvalue, that is,
\beq
{\cal H} \phi = E\phi,
\eeq
Again, we are omitting the index $\lambda$ in
the eigenfunctions and eigenvalues but
they contain the parameter $\lambda$ and thus 
depend on time through this parameter.
We wish to solve the Schrödinger equation with
the initial condition such that the wavefunction 
at $t=0$ is one of the eigenfunctions of the Hamiltonian.
Let us consider the following wave function
\beq
\psi_{\rm e} = e^{i\theta} \phi,
\label{85a}
\eeq
where $\theta$ is given by
\beq
\theta = -\frac{1}{\hbar}\int_0^t E dt',
\label{85}
\eeq
which is in accordance with the initial condition.
If we replace it in the Schrödinger equation, 
we see that it is not a solution because the
term $i\hbar\partial\phi/\partial t$
does not cancel out.
We assume then the following form for the solution
\beq
\psi = e^{i(\theta + u)}\phi,
\label{78}
\eeq
where $\theta$ is the dynamic phase given by (\ref{85})
and $u$ is a function to be found.
We look for a real solution for $u$ so that $\psi$
will differ from $\phi$ by a phase factor
which means that the system remains in the
state described by $\phi$.
In addition the wavefunction $\psi$ is normalized
because $\phi$ is normalized.

Replacing the expression (\ref{78}) in the Schrödinger
equation (\ref{60a}) we get the following equation
\beq
i\hbar \frac{\partial\phi}{\partial t}
 -\hbar\frac{\partial u}{\partial t}\phi
= e^{-i u}{\cal H}(e^{i u}\phi) - E\phi.
\label{82}
\eeq
Taking into account that ${\cal V}$ is a multiplying
operator, which is just a function,
the right hand side becomes
\beq
e^{-i u}{\cal K}(e^{i u}\phi) + {\cal V}\phi - E\phi,
\eeq
and the first term of this expression is
\beq
-\frac{\hbar^2}{2m} 
\left[i\frac{\partial^2 u}{\partial x^2} \phi
- \left(\frac{\partial u}{\partial x}\right)^2\phi
+ 2i\frac{\partial u}{\partial x}\frac{\partial\phi}{\partial x}
\right] + {\cal K}\phi.
\eeq
Replacing these results into equation (\ref{82})
we get
\[
i \frac{\partial\phi}{\partial t}
 -\frac{\partial u}{\partial t}\phi =
\]
\beq
= -\frac{\hbar}{2m} 
\left(i\frac{\partial^2 u}{\partial x^2} \phi
- (\frac{\partial u}{\partial x})^2\phi
+ 2i\frac{\partial u}{\partial x}\frac{\partial\phi}{\partial x}
\right),
\label{82a}
\eeq
which is an equation for $u$ as $\phi$ is known.

As we are looking for a real $u$, its imaginary
part should vanish and the real and imaginary parts of 
equation (\ref{82a}) become
\beq
-\frac{\partial u}{\partial t}
= \frac{\hbar}{2m} 
(\frac{\partial u}{\partial x})^2,
\label{82b}
\eeq
\beq
\frac{\partial\phi}{\partial t}
= -\frac{\hbar}{2m} 
\left(\frac{\partial^2 u}{\partial x^2} \phi
+ 2\frac{\partial u}{\partial x}\frac{\partial\phi}{\partial x}
\right),
\label{82c}
\eeq
where we have taken into account that $\phi$ is
real, a choice that is always possible to 
accomplish because ${\cal H}$ is Hermitian.
As we impose that the imaginary part of $u$
vanishes, this quantity should solve {\it both} 
the equations (\ref{82b}) and (\ref{82c}).

The first equation (\ref{82b}) can be solved by the separation
of variables. Assuming that $u(t,x)=\alpha(t) z(x)$ and
replacing it in (\ref{82}) we get
\beq
-\frac1{\alpha^2}\frac{\partial \alpha}{\partial t}
= \frac{\hbar}{2m} \frac1z 
(\frac{\partial z}{\partial x})^2.
\eeq
The left and right hand side should be a constant
that we choose to be equal to the unity, that is,
\beq
 -\frac{\partial \alpha}{\partial t} = \alpha^2,
\qquad\qquad
\frac{\hbar}{2m} (\frac{\partial z}{\partial x})^2 = z.
\eeq
Integrating,
\beq
\alpha = \frac{\varepsilon}{1+\varepsilon t},
\qquad\qquad
z = \frac{m}{2\hbar} x^2,
\label{84e}
\eeq
where $\varepsilon$ is a constant of integration.
Replacing these results in the second equation (\ref{82c}),
it becomes
\beq
2(1+\varepsilon t)\frac{\partial\phi}{\partial t}
= - \varepsilon \left(\phi + 2x\frac{\partial\phi}{\partial x}
\right).
\label{83}
\eeq

Using the scaling laws for $\phi$, we find
the equalities
\beq
\frac{x}{\phi}\frac{\partial\phi}{\partial x}
= \frac{\xi}{\hat\phi}\frac{\partial\hat\phi}{\partial\xi},
\eeq
\beq
\frac{1}{\phi}\frac{\partial\phi}{\partial t}
= \frac{a}{2\lambda}\left( 1 +
2 \frac{\xi}{\hat\phi}\frac{\partial\hat\phi}{\partial\xi}
\right) \frac{d\lambda}{dt}.
\eeq
Replacing these relations in equation (\ref{83})
we see that it becomes satisfied as long as 
\beq
\frac{d\lambda}{dt} = \frac{-\lambda\,\varepsilon}{a(1+\varepsilon t)}.
\label{86}
\eeq
The integration of this equation gives
\beq
\lambda = \lambda_0(1+\varepsilon t)^{-1/a},
\label{84}
\eeq
where $\lambda_0$ is a constant of integration,
which is the value of the parameter at $t=0$.
Therefore, $u$ solves {\it both} equation under
the condition (\ref{84}). In other words,
if the parameter $\lambda$ depends on time
in accordance with (\ref{84}), $u=\alpha z$ given by
(\ref{84}) is real and solves both equations
(\ref{82b}) and (\ref{82c}).

We may draw the following conclusions from the above
results. If the 
parameter $\lambda$ varies according to relation
(\ref{84}) and if the scaling (\ref{87a}) 
is fulfilled, then the
wave function given by equation (\ref{78})
is an exact solution and $u$ is real, that is,
it is indeed a phase, given by 
\beq
u = \frac{m\varepsilon x^2}{2\hbar(1+\varepsilon t)}.
\label{87}
\eeq
The solution (\ref{78}) is valid
for any value of the parameter $\varepsilon$,
and, of course, remains a solution when $\varepsilon$
is small, which characterizes a slow variation of
the parameter $\lambda$ because, according to
equation (\ref{86}) $d\lambda/dt$ is proportional
to $\varepsilon$.

\subsection{Harmonic oscillator of variable frequency}

The quantum harmonic oscillator with variable
frequency was treated by Husimi in 1953 \cite{husimi1953}.
The Hamiltonian is 
\beq
{\cal H} = - \frac{\hbar^2}{2m}\frac{\partial^2}
{\partial x^2} + \frac12m\omega^2 x^2,
\eeq
where the frequency $\omega$ depends on time.
The eigenfunctions $\phi_{k n}$ are
\cite{griffiths1994}
\beq
\phi_{n}(x) = \left(\frac{m\omega}{\pi\hbar}\right)^{1/4}
\frac1{\sqrt{2^n n!}} H_n(\xi)e^{-\xi^2/2},
\label{88}
\eeq
where $H_n(\xi)$ are the Hermite polynomials and
$\xi=x\sqrt{m\omega/\hbar}$. The corresponding
eigenvalues are
\beq
E_n = \hbar \omega(n+\frac12),
\eeq
where $n=0,1,2\ldots$
We wish to solve the Schrödinger equation
\beq
i\hbar \frac{\partial\psi}{\partial t} = {\cal H} \psi,
\eeq
with the initial condition that the initial state
is one of the eigenstates, say the eigenstate with
the quantum number $n$.

Writing the Hamiltonian in the form
\beq
{\cal H} = \hbar\omega (-\frac12 \frac{\partial^2}{\partial\xi^2}
+\frac12 \xi^2),
\eeq
it becomes manifest that it obeys the scaling relation 
(\ref{87a}) with $a=1/2$ and $b=1$ and $\omega$
playing the role of the parameter $\lambda$.
It is clear also that the eigenfunctions and eigenvalues
are in accordance with the scaling relations (\ref{87b}) and
(\ref{87c}). 

From the results obtained above, a closed solution for
the time dependent Schrödinger equation 
can be obtained for the following time dependence
of the frequency
\beq
\omega = \frac{\omega_0}{(1+\varepsilon t)^2}.
\eeq
The solution is
\beq
\psi = e^{i(\theta + u)} \phi_n,
\eeq
where $\phi_n$ is one of the eigenfunctions and
\beq
\theta
= - \omega_0(n+\frac12)
(t + \varepsilon t^2 + \frac13\varepsilon^2 t^3),
\eeq
and
\beq
u = \frac{\varepsilon \,m x^2}{2\hbar(1+\varepsilon t)}.
\eeq

\subsection{Raising and lowering operators}

We solve again the harmonic oscillator but now
we use a representation in terms of the 
lowering and raising operators defined by
\beq
a = \sqrt{\frac{m\omega}{2\hbar}} x
+ i \frac{p}{\sqrt{2m\hbar\omega}},
\label{90a}
\eeq
\beq
a^\dagger = \sqrt{\frac{m\omega}{2\hbar}} x
- i \frac{p}{\sqrt{2m\hbar\omega}},
\label{90b}
\eeq
where $p=-\hbar\partial/\partial x$ is the momentum
operator.
They hold the relations $a|n\ra=\sqrt{n}|n\ra$
and $a^\dagger|n\ra=\sqrt{n+1}|n\ra$, 
and fulfills the commutation relation $[a,a^\dagger]=1$.

In terms of these lowering and raising operators,
the Hamiltonian 
\beq
H = \frac{p^2}{2m} + \frac12 m\omega^2 x^2
\eeq
of the harmonic oscillator becomes
\beq
H = \hbar\omega (a^\dagger a+\frac12).
\eeq
The eigenvectors of $H$ are $|n\ra$,
that is,
\beq
H|n\ra = E_n | n\ra,
\eeq
and the eigenvalues are $E_n=\hbar\omega(n+1/2)$.

We wish to solve the Schrödinger equation
\beq
i\hbar \frac{d}{dt}|\psi\ra = H|\psi\ra,
\eeq
considering that the frequency $\omega$ is a
time dependent parameter and that at $t=0$
the oscillator is in one of its eigenstates.

As $\omega$ is a time dependent parameter
the operators $a$, $a^\dagger$, and the
state vectors $|n\ra$ depend on time
through $\omega$. It is thus convenient to
determine their variation with $\omega$.
From the definitions given by (\ref{90a}) and
(\ref{90b}), it follows that $a$ and $a^\dagger$
vary with $\omega$ according to 
\beq
\frac{\partial a}{\partial\omega} = \frac1{2\omega} a^\dagger,
\qquad\qquad \frac{\partial a^\dagger}{\partial\omega} = \frac1{2\omega} a.
\eeq
From these relations we find
\beq
\frac{dH}{d\omega} = \frac{\partial H}{\partial\omega}
= m\omega x^2.
\eeq
From (\ref{88}) we establish the following variation of
the eigenvectors with $\omega$,
\beq
\frac{d}{d\omega}|n\ra
= \frac1{4\omega}(aa-a^\dagger a^\dagger)|n\ra
= \frac{i}{4\hbar\omega}(xp+px)|n\ra.
\label{91}
\eeq

To find the solution of the Schrödinger equation we
consider the following state vector 
\beq
|\psi\ra = e^{i(\theta + \alpha z)}|n\ra,
\label{46}
\eeq
where 
\beq
\theta = -\frac1\hbar \int_0^t E_n dt',
\eeq
and $\alpha$ is a time dependent scalar, and $z=(m/2\hbar)x^2$,
$x^2$ being the position operator squared.
Replacing $|\psi\ra$ into the Schrödinger equation, we find
\beq
-\frac{d\alpha}{dt} \frac{m}{2}x^2|n\ra
+ i\hbar \frac{d}{dt} |n\ra
= e^{-i \alpha z}H e^{i \alpha z} |n\ra - E_n |n\ra.
\eeq

Now, recalling that $z$ is proportional to $x^2$,
\[
e^{-i \alpha z}H e^{i \alpha z} = \frac1{2m}
e^{-i \alpha z} p^2 e^{i \alpha z} + \frac12m\omega^2 x^2 =
\]
\beq
= H + \frac{\alpha}{2}
(xp + px) + \frac{\alpha^2 m}{2} x^2,
\eeq
and the Schrödinger equation becomes
\[
- \frac{d\alpha}{dt}\frac{m}{2} x^2|n\ra
+ i \hbar\frac{d\omega}{dt}\frac{d}{d\omega} |n\ra =
\]
\beq
= \frac{\alpha}{2}(xp + px)|n\ra
+ \frac{\alpha^2 m}{2} x^2|n\ra.
\eeq
A solution of this equation occurs if
\beq
\frac{d\alpha}{dt} = - \alpha^2,
\eeq
from which follows the result 
\beq
\alpha = \frac{\varepsilon}{1+\varepsilon t},
\eeq
and if
\beq
\frac{d\omega}{dt} = - 2\alpha\omega,
\eeq
where we have taken into account the
relation(\ref{91}). Solving this equation 
one finds the dependence of $\omega$ with time,
\beq
\omega = \frac{\omega_0}{(1+\varepsilon t)^2}.
\label{46a}
\eeq
We conclude that (\ref{46}) is an exact solution
if $\omega$ depends on time according to (\ref{46a}).

\section{Statistical mechanics}

\subsection{Hertz invariant}

Let ${\cal H}(q,p,\lambda)$ be the Hamiltonian of
a system with $n$ degrees of freedom where we are denoting
by $q$ and $p$ the collection of coordinates
$q_1,\ldots,q_n$ and momenta $p_1,\ldots,p_n$, and
$\lambda$ is a parameter.
In his treatise on statistical mechanics \cite{gibbs1902}, 
Gibbs introduced the following integral
\beq
\Phi = \int_{{\cal H}\leq E} dq dp,
\label{22}
\eeq
which is the hypervolume of the region of the phase space 
enclosed by a hypersurface of constant energy $E$.
He also defines the quantity
\beq
\Omega = \frac{\partial\Phi}{\partial E}.
\label{22c}
\eeq

Using the step function $\vartheta(x)$, which
is equal to zero or the unity according to whether
the argument $x$ is negative or positive,
then $\Phi$ can be written as
\beq
\Phi = \int \vartheta(E - {\cal H}(\lambda)) dqdp.
\label{22a}
\eeq
Taking into account that the derivative of the
step function is the Dirac delta function $\delta(x)$,
then $\Omega$ can be written as
\beq
\Omega = \int \delta(E - {\cal H}(\lambda)) dq dp.
\label{22b}
\eeq

Gibbs considers two possible forms for the
entropy of a system. One of them is
\beq
S = k\ln\Omega,
\label{111a}
\eeq
which is the form widely used, and the other is
\beq
S' = k\ln\Phi.
\label{111b}
\eeq
He states that each of them has its
advantage but the first form is a little more
simple than the second and if simplicity
is a criterion, the first form is preferable.
Nevertheless, the two forms differ little
from one another when the number of degrees
of freedom is large, a feature that was recognized
by Gibbs \cite{gibbs1902}.
Another appeal for the use of the first form is that 
$\Omega$ occurs as the normalization of the 
Gibbs microcanonical distribution \cite{gibbs1902},
given by
\beq
\rho = \frac1{\Omega}\delta(E-{\cal H(\lambda})).
\label{25a}
\eeq

In accordance with Clausius, who introduced the concept
of thermodynamic entropy, this quantity is constant along
a reversible adiabatic process, understood as a process
carried out without the exchange of heat, and slow enough
so that the system can be considered to be in equilibrium.
If $S$ and $S'$ are to be interpreted as the thermodynamic
entropy, then $\Phi$ and $\Omega$ should be constant
along a reversible adiabatic process. The question is thus
how to define a mechanical procedure that results in a
reversible adiabatic process, without
referring to heat. This question was implicit answered by
Paul Hertz in a paper of 1910 \cite{hertz1910}. In this paper
he used a procedure that coincides with what we are calling
slow parametric action to show that $\Phi$ is constant. 
He assumed that the system has an external coordinate
that is changed by external intervention, and that the
energy is a function of $q$, $p$ and $a$. The variables
$q$ and $p$ vary according to the equations of motion
and the parameter $a$ is subject to our arbitrariness.

The approach used by Hertz was based on a hint
contained in the Gibbs treatise on
statistical mechanics \cite{jammer1966}. 
According to Gibbs \cite{gibbs1902} ``the entropy of
a body is not (sensibly) affected by mechanical action,
during which the body is at each instant (sensibly)
in a state of thermodynamic equilibrium'', which
``may usually be attained by a sufficiently slow
variation of the external coordinates''.
The slow variation of the parameter was implicit
in the Hertz use of the Gibbs microcanonical distribution.
If the variation of the parameter is slow, the system 
remains in equilibrium and the Gibbs distribution, which is
understood to be valid for system in equilibrium,
can be used. Therefore, we may say that {\it the use of the
Gibbs distribution for two distinct values of the parameter,
resulting from its variation with time, means that the
variation is implicitly slow}. 
The invariance of $\Phi$ is thus a parametric invariance,
although Hertz did not used this terminology but simply
stated that $\Phi$ is constant.

The demonstration of the invariance of $\Phi$ is
contained in some books on statistical mechanics
\cite{becker1967,munster1969,toda1983} and
has been considered by some authors
\cite{fernandez1982,rugh2001,dunkel2006,campisi2008,uline2008}.
A demonstration based on the Hertz paper was
provided by de Koning and Antonelli \cite{dekoning1996}. 
We demonstrate the invariance of $\Phi$ as follows. 
The variation in energy of a
system described by a time dependent Hamiltonian is
\beq
\frac{d{\cal H}}{dt} = \frac{\partial{\cal H}}{\partial t}.
\eeq
If the Hamiltonian depends on time through a parameter 
$\lambda$ which varies slowly with time, then
in accordance with (\ref{31}), the right-hand side
of this equation is replaced by its time average and
it becomes
\beq
\frac{dE}{d\lambda} = 
\overline{\frac{\partial{\cal H}}{\partial\lambda}},
\eeq
where $E=\overline{\cal H}$. The time average is then
replaced by the average in probability,
\beq
\frac{dE}{d\lambda}
= \la \frac{\partial{\cal H}}{\partial\lambda} \ra,
\label{93}
\eeq
\beq
\la \frac{\partial{\cal H}}{\partial\lambda} \ra 
= \int \frac{\partial{\cal H}}{\partial\lambda}\rho\, dqdp,
\eeq
and $\rho$ is the Gibbs microcanonical distribution, and
\beq
E = \la{\cal H}\ra = \int {\cal H}\rho dqdp.
\eeq

The invariance of $\Phi$ means
that $\Phi(E',\lambda')$ equals $\Phi(E,\lambda)$
when the parameter changes slowly from $\lambda$
to $\lambda'$ causing a change of energy from $E$ 
to $E'$. Considering small differences, we write  
\beq
\frac{d\Phi}{d\lambda} = \frac{\partial\Phi}{\partial\lambda}
+ \frac{\partial\Phi}{\partial E}\frac{dE}{d\lambda},
\label{24}
\eeq
and the invariance means that $d\Phi/d\lambda$ should
vanish.
Using the relation (\ref{22c}), the equation (\ref{24}) can be 
written as
\beq
\frac{d\Phi}{d\lambda} = \frac{\partial\Phi}{\partial\lambda}
+ \Omega\frac{dE}{d\lambda}.
\label{24a}
\eeq

We observe that, using the definition (\ref{25a})
of the microcanonical probability  distribution,
the equation (\ref{93}) can be written as
\beq
\Omega \frac{dE}{d\lambda} = \int \frac{\partial{\cal H}}{\partial\lambda}
\delta(E-{\cal H}) dqdp.
\label{93a}
\eeq
We also observe that, using the expression (\ref{22a}),
the derivative of $\Phi$ with respect to $\lambda$ is
\beq
\frac{\partial\Phi}{\partial\lambda}
= - \int \frac{\partial{\cal H}}{\partial\lambda}
\delta(E - {\cal H}) dq dp.
\label{23}
\eeq
From these two relations, it becomes manifest that the
right hand side of (\ref{24a}) vanishes and so does
$d\Phi/d\lambda$ as desired.

Let us calculate $\Phi$ for a system 
of $N$ particles confined in a container of
volume $V$. The number of degrees of freedom is $3N$.
The particles do not interact
and the energy of the system is the kinetic energy
\beq
{\cal H} = \sum_i \frac{p_i^2}{2m},
\eeq
where $p_i$ denotes a component of each one of the
particles. 
The quantity $\Phi$ is 
\beq
\Phi = V^N\int_{{\cal H}\leq E} d^{3N} p,
\eeq
the integral being equal to the volume of a sphere of radius
$\sqrt{2mE}$ in a space of dimension $3N$, that is,
\beq
\Phi = V^N \frac{(2\pi mE)^{3N/2}}{(3N/2)!},
\label{43}
\eeq
and
\beq
\Omega = \frac{3N}{2E}\Phi.
\eeq

\subsection{Several parameters}

Let us generalize the above result for several 
parameters that we denote by $\lambda_i$.
The variation in time of the Hamiltonian is
\beq
\frac{d{\cal H}}{dt} = \sum_i \frac{\partial{\cal H}}{\partial\lambda_i}
\frac{d\lambda_i}{dt}.
\eeq
Considering that the parameter vary slowly in time this equation is
replaced by
\beq
dE = \sum_i \overline{\frac{\partial{\cal H}}{\partial\lambda_i}}
d\lambda_i.
\eeq
The time averages are in turn replaced by averages on probability,
\beq
dE = \sum_i \la{\frac{\partial{\cal H}}{\partial\lambda_i}}\ra
d\lambda_i.
\label{121}
\eeq
We write this equation in the form
\beq
dE = - \sum_i F_i d\lambda_i,
\label{102}
\eeq
where
\beq
F_i = - \la{\frac{\partial{\cal H}}{\partial\lambda_i}}\ra
= - \int\frac{\partial{\cal H}}{\partial\lambda_i}\rho dqdp,
\eeq
and $\rho$ is the Gibbs microcanonical distribution
\beq
\rho = \frac1\Omega \delta(E-{\cal H}),
\eeq
where
\beq
\Omega = \int \delta(E-{\cal H}) dqdp,
\eeq
and depends on $\lambda_i$. Defining the integral
\beq
\Phi = \int\vartheta(E-{\cal H})dqdp,
\label{107}
\eeq
that depends on $\lambda_i$, we see that $F_i$
can be written as
\beq
F_i = \frac1\Omega \frac{\partial\Phi}{\partial\lambda_i}.
\label{107a}
\eeq
Replacing this results in equation (\ref{102}),
and taking into account that
\beq
\Omega = \frac{\partial\Phi}{\partial E},
\eeq
we find
\beq
\frac{\partial\Phi}{\partial E} dE
+ \sum_i \frac{\partial\Phi}{\partial\lambda_i} d\lambda_i = 0.
\eeq
But the left-hand side of this equation is the differential
$d\Phi$. Since it equals zero, it follows that $\Phi$
is constant as we vary $E$ and the parameters $\lambda_i$.

\subsection{Variable number of particles}

The integral $\Phi$ given by the equation (\ref{22}) was
shown to be an invariant when a parameter changes slowly. 
However, the variation of the parameter did not change
the number of particles. Here wish to show that 
the expression that is invariant when the number $n$ of
particles changes is $\Phi_n=\Phi/n!$ with the factor $1/n!$.

Let ${\cal H}_n$ be the Hamiltonian function corresponding
to a system of $n$ particles. We use the notation
$x_i$ for the positions and momenta of the $i$-th particle
so that the 
Hamiltonian ${\cal H}_n$ depends on the variables from $x_1$
to $x_n$ and is invariant under the permutation of the $n$ particles.

We now consider the removal of a particle from the system.
To simulate this procedure we suppose that at $t=0$ 
a particle is chosen at random and it its velocity is
slowly reduced and its interaction with the
other particles is slowly decreased. After an interval
of time $\Delta t$, its velocity has vanished and it
does not interact with other particles anymore.
We suppose that this procedure is carried out
by means of a parameter $\lambda$ that takes the value 
$\lambda_0$ at $t=0$ and the value $\lambda_1$ at
$t=\Delta t$. During this procedure the Hamiltonian
${\cal H}$ that describes the system is not invariant
by permutation of all the particles. If we
chose the particle $n$ to be taken out, ${\cal H}$
is invariant under the permutation of the remaining
$n-1$ particles.

Supposing that $\lambda$ varies slowly, then the variation
of the energy $E$ with $\lambda$ is given by
\beq
\frac{dE}{d\lambda} = \la \frac{\partial{\cal H}}{\partial\lambda}\ra,
\label{32}
\eeq
where the average is calculated using the probability
distribution
\beq
\rho = \frac1{\Omega}\delta(E-{\cal H}),
\label{25b}
\eeq
where
\beq
\Omega = \int \delta(E-{\cal H})\,d^n x,
\eeq
and depends on $E$ and $\lambda$.
It is convenient to define the integral 
\beq
\Phi = \int \vartheta(E - {\cal H})\,d^n x,
\label{22d}
\eeq
so that
\beq
\frac{\partial\Phi}{\partial E} = \Omega.
\label{32a}
\eeq

Deriving (\ref{22d}) with respect to $\lambda$, and taking into
account that ${\cal H}_n$ depends on $\lambda$,
we find
\beq
\frac{\partial\Phi}{\partial\lambda}
= - \int \frac{\partial{\cal H}}{\partial\lambda}
\delta(E - {\cal H})\,d^n x,
\label{22e}
\eeq
from which we obtain after dividing by $\Omega$
\beq
\frac{dE}{d\lambda} = - \frac1{\Omega}
\frac{\partial\Phi}{\partial\lambda}.
\eeq

Replacing (\ref{32a}) into this equation, we
reach the following relation
\beq
\frac{dE}{d\lambda}
= - \frac{\partial\Phi/\partial\lambda}{\partial\Phi/\partial E}
= \left(\frac{\partial E}{\partial\lambda}\right)_\Phi.
\eeq
This relation tell us that $E$ and $\lambda$ varies in 
such a way that $\Phi(E,\lambda)$ is invariant.

If $E$ is the energy when $\lambda=\lambda_0$ and $E'$ when
$\lambda=\lambda_1$ then $\Phi(E,\lambda_0)=\Phi(E',\lambda_1)$.
Let us define $\Phi_n^*$ by
\beq
\Phi_n^* = \int\vartheta(E-{\cal H}_n)d^n x
\eeq
then $\Phi(E,\lambda_0)=\Phi_n^*(E)$ and
$\Phi(E',\lambda_1)=n\Phi_{n-1}^*(E')$, 
the factor $n$ coming from the existence of
$n$ possibility of removing a particle from the
system, and the invariance is written as
\beq
\Phi_n^*(E) = n \Phi_{n-1}^*(E')
\eeq
Defining $\Phi_n=\Phi_n^*/n!$ then the invariance
becomes
\beq
\Phi_n(E) = \Phi_{n-1}(E')
\eeq
From these relation it follows that
\beq
\Phi_n(E) = \frac1{n!}\int \vartheta(E - {\cal H}_n)\,d^n x
\label{106}
\eeq
is invariant when one varies $E$ and $n$.

For a system of $N$ noninteracting particles
we use the result (\ref{43}) to find 
\beq
\Phi_N = \frac{V^N}{N!} \frac{(2\pi mE)^{3N/2}}{(3N/2)!}.
\label{43a}
\eeq

\subsection{Quantum systems}

Here we use the occupation number representation
to describe the dynamics of a quantum system
\cite{fetter1971}.
To this end we introduce the operator $a_i^\dagger$
which creates a particle in the single particle
state $\psi_i$ and the operator $a_i$ which 
annihilates a particle in state $\psi_i$. 
The number operator is $a_i^\dagger a_i$
and the total number operator is
\beq
N = \sum_i a_i^\dagger a_i.
\eeq

We assume the following form for the Hamiltonian
\beq
H = \sum_{ij} K_{ij} a_i^\dagger a_j +
\sum_{ijlk} V_{ijkl} a_i^\dagger a_j^\dagger a_l a_k,
\eeq
where $K_{ij}=\la \phi_i|K|\phi_j\ra$ and
$V_{ijkl}=\la \phi_i\phi_j|V|\phi_k\phi_l\ra$,
and $K$ and $V$ are the kinetic and potential
energy. We remark that $N$ commutes with $H$.
If $H$ does not depend explicitly on time,
$H$ is conserved and so is $N$.

Now we wish to treat the case where the number
of particle varies with time, for instance, by
introducing or removing particles from the system.
This is accomplished by assuming that the creation
and annihilation operators depends on a parameter
$\lambda$ that varies with time. We start with
the equation
\beq
\frac{dE}{d\lambda} = {\rm Tr} \frac{\partial H}{\partial\lambda} \rho,
\eeq
where $\rho$ is the density operator, which we assume
to be given by
\beq
\rho = \frac1{\Omega} \delta(E - H),
\eeq
and
\beq
\Omega = {\rm Tr} \delta(E - H).
\eeq
In an explicitly form,
\beq
\frac{dE}{d\lambda} = \frac1\Omega
{\rm Tr} \frac{\partial H}{\partial\lambda} \delta(E-H).
\eeq

Defining 
\beq
\Phi = {\rm Tr} \vartheta(E-H),
\eeq
we find the relations
\beq
\frac{\partial\Phi}{\partial E} = \Omega,
\eeq
\beq
\frac{\partial\Phi}{\partial\lambda}
= - {\rm Tr} \frac{\partial H}{\partial\lambda} \delta(E-H),
\eeq
from which follows
\beq
\frac{dE}{d\lambda}
= - \frac{\partial\Phi/\partial\lambda}{\partial\Phi/\partial E}
= \left(\frac{\partial E}{\partial\lambda}\right)_\Phi.
\eeq
Therefore as one varies $E$ and $\lambda$, $\Phi$ remains invariant.

Let us suppose that as $\lambda$ varies, assuming values
$\lambda_1$, $\lambda_2$, and so on, the number of particles 
in the system are, respectively, $n_1$, $n_2$, and so on,
and the energy are $E_1$, $E_2$, and so on. If
we  define
\beq
\Phi_n^*(E) = {\rm Tr}_n \vartheta (E-H),
\label{105}
\eeq
where the trace is taken within the subspace of states with
$n$ particles, then $\Phi(E_i,\lambda_i)=\Phi_{n_i}(E_i)$.
Therefore the invariance $\Phi(E,\lambda)$ when one varies
$E$ and $\lambda$ is equivalent to the invariance of
$\Phi_n^*(E)$ when one varies the energy $E$ and the number
of particles. 

The interpretation of $\Phi_n^*(E)$ is very simple.
If we use a basis of eigenvectors $|\phi_i^{(n)}\ra$ of $H$
belonging to the Hilbert space with $n$ particles,
with the associate eigenvalue $E_i^{(n)}$, then
\beq
\Phi_n^* = \sum_i \vartheta(E-E_i^{(n)}).
\eeq
Thus the invariant $\Phi_n^*$ is the number of quantum
states with $n$ particles with eigenvectors less or equal
to $E$.

It is worth taking the classical limit of (\ref{105}).
To this end, we consider that the system is enclosed in a cubic 
vessel of size $L$ and use the eigenfunctions of the 
momentum
\beq
\phi_n(x) = \frac1{\sqrt{L}} e^{ikx}, 
\eeq
where $k=2\pi n$. After expressing $\Phi$ in terms of the
basis of plane waves, we may take the classical limit. The result is
\beq
\Phi_N^* = \frac1{N!}\int \vartheta(E-{\cal H}) \frac{d^nx d^np}{h^n},
\eeq
where ${\cal H}$ is the classical Hamiltonian,
$N$ is the number of particles, $n$ is the 
number of degrees of freedom, and $h$ is the Planck constant.
Except for the factor $h^n$, this coincides with (\ref{106}).

For a system of $N$ noninteracting particles
we use the result (\ref{43a}) to find 
\beq
\Phi_N^* = \frac{V^N}{N!} \frac{(2\pi mE/h^2)^{3N/2}}{(3N/2)!}.
\label{43b}
\eeq

\subsection{Canonical distribution}

Let us suppose that instead of the Gibbs microcanonical
distribution the system is described by the 
Gibbs canonical distribution 
\beq
\rho = \frac1Z e^{-\beta{\cal H}},
\label{125}
\eeq
where $\beta$ is a parameter, ${\cal H}$ is the
Hamiltonian that does not depend on $\beta$
but depend on a parameter $\lambda$, and 
\beq
Z = \int e^{-\beta{\cal H}} dqdp
\eeq
We wish to determine the invariant which is
analogous to that of Hertz.

The probability distribution $P(E)$ of the energies is the 
marginal probability distribution obtained
from $\rho(q,p)$ and is given by
\beq
P(E) = \frac1Z \Omega e^{-\beta E},
\eeq
where
\beq
\Omega = \int \delta(E-{\cal H})dqdp,
\eeq
which depends on $E$ and $\lambda$, and
\beq
Z = \int e^{-\beta E} \Omega dE,
\label{126}
\eeq
which depends on $\beta$ and $\lambda$.

If the parameter $\lambda$ is varying slowly in time we use equation
(\ref{93})
\beq
\frac{dU}{d\lambda} = \la{\frac{\partial{\cal H}}{\partial\lambda}}\ra,
\label{121a}
\eeq
but now the average on the right-hand side is taken
over the canonical distribution $\rho$,
\beq
\la{\frac{\partial{\cal H}}{\partial\lambda}}\ra = 
\int \frac{\partial\cal{H}}{\partial\lambda}\rho dqdp.
\label{121b}
\eeq
The equation (\ref{121a}) determines the relation between 
$\beta$ and $\lambda$ along a parametric slow action.
The invariant $\Psi$ that we wish to find will then
be a function of $\beta$ and $\lambda$.

The right-hand side of (\ref{121b}) is written in a more
convenient form as
\beq
\frac1Z\int \frac{\partial\cal{H}}{\partial\lambda}
\delta (E-{\cal H}) e^{-\beta E} dqdp dE,
\eeq
which is equal to 
\beq
- \frac1Z\int \frac{\partial\Phi}{\partial\lambda} e^{-\beta E} dE,
\eeq
where
\beq
\Phi = \int \vartheta (E-{\cal H}) dqdp,
\eeq
and depends on $E$ and $\lambda$.

The average energy $U$ is written as
\beq
U = \frac1Z \int E e^{-\beta E} \Omega dE, 
\eeq
and can be obtained from $Z$ by
\beq
U = - \frac1Z \frac{\partial Z}{\partial\beta}.
\label{127}
\eeq

Using the above results, the equation (\ref{121a})
becomes
\beq
\frac{dU}{d\lambda}
= - \frac1Z \int \frac{\partial\Phi}{\partial\lambda}e^{-\beta E} dE.
\label{124}
\eeq

The invariant $\Psi$,
which depends on $\beta$ and $\lambda$, is such
that if $\Psi(\beta,\lambda)$ is held constant then
$\beta$ is connected to $\lambda$ in such a way that
(\ref{124}) is fulfilled. It is equivalent to say
that, in equation (\ref{124}),
\beq
\frac{d\beta}{d\lambda} = - \frac{\partial\Psi/\partial\lambda}
{\partial\Psi/\partial\beta}.
\eeq
For comparison with equation (\ref{124}), 
we write this equation in the form
\beq
\frac{\partial\Psi}{\partial\beta}
\frac{d\beta}{d\lambda} = - \frac{\partial\Psi}{\partial\lambda}.
\label{128}
\eeq

We show now that the following form,
\beq
\Psi = e^{\beta U} Z,
\label{45a}
\eeq
is invariant, where $Z$ is given by (\ref{126}),
and is equivalent to the expression
\beq
Z = \beta \int \Phi e^{-\beta E} dE,
\label{45}
\eeq
obtained by an integration by parts. 
To this end we calculated its
derivative with respect to $\beta$ and $\lambda$.
Recalling that $\Phi$ does not depend on $\beta$, 
and using the result (\ref{127}), we find
\beq
\frac{\partial\Psi}{\partial\beta}
= \beta \frac{\partial U}{\partial\beta} \Psi.
\eeq
The other derivative is
\beq
\frac{\partial\Psi}{\partial\lambda}
= \left(\beta\frac{\partial U}{\partial\lambda}
+ \frac{\beta}Z \int\frac{\partial\Phi}{\partial\lambda}
e^{-\beta E} dE\right)\Psi.
\eeq

Replacing the two derivatives in equation (\ref{128}), we get
\beq
\frac{\partial U}{\partial\beta}
\frac{d\beta}{d\lambda} = 
- \frac{\partial U}{\partial\lambda}
- \frac1Z \int\frac{\partial\Phi}{\partial\lambda}
e^{-\beta E} dE,
\eeq
which is equivalent to equation (\ref{124}),
and shows that $\Psi$ is an invariant.

For a system of $N$ noninteracting particles
we replace the result (\ref{43b}) in equation (\ref{45})
to find, after integration in $E$,
\beq
Z = \frac{V^N}{N!} (\frac{2\pi m}{\beta h^2})^{3N/2}.
\label{45b}
\eeq
Taking into account that $U=3N/2\beta$ we obtain the 
invariant $\Psi$,
\beq
\Psi = e^{3N/2}\frac{V^N}{N!} \left(\frac{2\pi m}{\beta h^2}\right)^{3N/2}.
\eeq

\section{Thermodynamics}

\subsection{Adiabatic invariants}

An adiabatic process is understood as a process 
which does not involve the exchange of heat. 
This type of process played an important role in the
development of the theory of heat
\cite{oliveira2018,oliveira2019}. 
Laplace for instance
proposed that the variations of pressure in the
propagation of sound in air are adiabatic
processes. Carnot announced his fundamental principle
by means of a cyclic process involving two isothermal
and two adiabatic subprocesses. Such a cyclic process was
also used by Clausius when he introduced the laws of
thermodynamics. The term adiabatic was coined by Rankine
in 1859 to refer to processes without the intervention of heat.

An adiabatic process may be a slow process or
a very fast one as long as the system 
is enclosed by adiabatic walls or is in isolation.
If the walls are not perfectly adiabatic, the process
can still be adiabatic if it 
occurs very rapidly. If an air pump is 
compressed quickly, there is no time for heat
to be exchanged with the surroundings and the 
process is adiabatic. In this case, the temperature
of the air inside the pump increases. Just after
compression, considering that the walls of the
pump are not perfectly adiabatic, heat is 
release to the surrounding, which is
felt by the pump handler. Thus, neither the 
slowness nor the quickness are distinguishing
features of an adiabatic process.

Along an adiabatic process we may ask whether there 
are state functions that are invariant along this
process. If the process is slow enough, this question
was answered by Clausius when he introduced in 1854
a quantity, which in 1865 he called entropy
that remains constant along a slow adiabatic process
\cite{clausius1854,clausius1865,clausius1867}.
But before Clausius, it was known that the quantity
$pV^\gamma$, remains constant when an ideal gas
undergoes a slow adiabatic process.
Here, $p$ is the pressure, $V$ is the volume and
$\gamma$ is the ratio of the two types of specific heats.

The result that $pV^\gamma$ is a constant was obtained
theoretically by Poisson in 1823 \cite{poisson1823}
by using the equation of state of an ideal gas, $pV$
proportional to the temperature $T$, and the assumption
that $\gamma$ is constant. The starting point of his
derivation is the equation
\beq
\frac{\gamma dV}{\partial V/\partial T}
+ \frac{dp}{\partial p/\partial T} = 0,
\label{4}
\eeq
valid along an adiabatic process,
which for an ideal gas becomes $pdV + Vdp$.
Considering that $\gamma$ is constant, the integration
of this equation gives $pV^\gamma$ equal to a constant.
Although the equation (\ref{4}) was derived by Poisson
by assuming that heat is a state function, nevertheless
it remains valid within thermodynamics.
In fact, the invariance of $pV^\gamma$ was derived
by Clausius in 1850 within the realm of thermodynamics 
\cite{clausius1867,clausius1850}.

The derivation of the
adiabatic invariant $pV^{\gamma}$, either by Poisson
or by Clausius, involved thermal properties 
that depended explicitly on the temperature.  
However, as an adiabatic process involves no heat we may
presume that any adiabatic invariant could be derived
without referring to thermal properties. That is, a
derivation carried out within the realm of mechanics by
considering a parametric slow process. 
To show that this is indeed
the case for the Poisson adiabatic invariant, we follow
the reasoning of Rayleigh, which is summarized by equation
(\ref{34}).

If the volume $V$ of a gas enclosed in a vessel
is varied slowly, then in accordance with the equation
(\ref{34}), the variation of the energy
with the volume  is
\beq
\frac{dE}{dV}= - p,
\label{34a}
\eeq 
where $p$ is the pressure of the gas.
For a system of noninteracting molecules,
it follows from the laws of mechanics that
the pressure $p$ is two-thirds of the kinetic
energy $E$,
\beq
p = \frac{2E}{3V},
\label{34b}
\eeq
a result derived by Krönig \cite{kronig1856}, by Clausius
\cite{clausius1857,clausius1857a}, and by Maxwell \cite{maxwell1860}
within the kinetic theory. Considering
a simple gas with only translational degrees of
freedom, $E$ is the total energy. 
Replacing the result (\ref{34b}) into (\ref{34a}),
we find by integration that
$EV^{2/3}$ is constant, from which 
follows that $p V^{5/3}$ is an adiabatic invariant.

A similar invariant is obtained for the radiation.
In his treatise on electricity and magnetism
of 1873 \cite{maxwell1873}, Maxwell showed that the pressure
of radiation is one-third of the density of energy, or 
\beq
p = \frac{E}{3V}.
\label{42}
\eeq
Replacing this result into (\ref{34a}), we find
by integration that $EV^{1/3}$ from which follows
that $pV^{4/3}$ is an adiabatic invariant.

It is worth mentioning that in the derivation of
the Stefan-Boltzmann law carried out by Boltzmann
in a paper of 1884 \cite{boltzmann1884},
he made use of the Maxwell relation (\ref{42})
along with thermodynamic reasoning.
A derivation of this law by the use of the 
invariant just obtained is as follows.
As the entropy $S$ is an invariant we may
consider it as a function of the invariant
$EV^{1/3}$. Assuming that it is a
homogeneous function of $E$ and $V$, it
follows that $S$ is proportional to $V^{1/4}E^{3/4}$.
The temperature is obtained by $1/T=\partial S/\partial E$
from which follows that $E=aVT^4$. Replacing this
result in equation (\ref{42}), we find $p$ to be
proportional to $T^4$, which is a statement of the
Stefan-Boltzmann law. 

\subsection{Boltzmann and Gibbs entropy}

The concept of entropy was introduced by Clausius
as a quantity that is constant along a slow adiabatic
process and that increases in an irreversible process.
These properties of the entropy are a brief statement
of the second law of thermodynamics introduced by Clausius.
The definition of entropy that emerges from the
Boltzmann writings is laid down on the second 
property, related to the increase of entropy,
and can be found in his book on the theory of gases
\cite{boltzmann1896,boltzmann1964}.

In a paper of 1872 Boltzmann proved that the
quantity \cite{boltzmann1872} 
\beq
H = \int f \ln f d^3rd^3v
\label{133}
\eeq
never increases, where $f$ is the one-particle probability
distribution. The negative of
$H$ was understood by Boltzmann as proportional to the entropy.
To demonstrate this result, which comprises the Boltzmann
H-theorem, Boltzmann used the transport equation that he
introduced in the same paper. Replacing the Maxwell
distribution in the expression for $H$, he obtained the
following expression for the entropy of an ideal gas
\cite{boltzmann1872}
\beq
N \ln V \left(\frac{4\pi T}{3m}\right)^{3/2}\!\! + \frac32 N,
\eeq
where $N$ is the number of molecules, $V$ the volume,
$m$ the mass of a molecule, and $T$ is the mean
kinetic energy.

Later on, in 1877,
Boltzmann related entropy to probability, which he stated
in the following terms \cite{boltzmann1877,dugas1959}.
{\it In most cases the initial state of a system
is a very improbable one and the system has the tendency
to reach more probable states, those of thermal equilibrium. 
If we apply this to the second law, we can identify that
quantity, which is usually called entropy, with the probability
of the state in question}.
For an ideal gas of $n$ molecules, Boltzmann finds the
relation between entropy and probability as follows
\cite{boltzmann1877,dugas1959}. The number of complexions
that corresponds to a given repartition 
$n_0$, $n_1$, $n_2$, and so on, of the $n$ molecules into
the kinetic energies $0$, $\varepsilon$, $2\varepsilon$, and 
so on, is the permutation number
\beq
P = \frac{n!}{n_0!n_1!n_2!\ldots}.
\eeq
The most probable state corresponds to the maximum of $P$,
or equivalently to the maximum of $\ln P$. Considering that
$n_i$ are large numbers we may use the approximation
$\ln n! = n\ln n -n$ to find
\beq
\ln P = n \ln n - \sum_i n_i\ln n_i.
\eeq
Taking the continuous limit of the energy, the second
term of this expression yields the quantity $-H$,
which is identified as the entropy. 
Boltzmann formulates a general principle relating
$P$ and entropy in following terms
\cite{boltzmann1877,dugas1959}.
{\it The measure of permutability of all bodies will
always grow in the course of the state changes and can
at most remain constant as long as all bodies are in
thermal equilibrium}.

In a paper of 1901 on the radiation formula,
Planck translated the Boltzmann 
relation between entropy and probability in the
following terms \cite{planck1901}
\beq
S = k \ln W,
\label{110}
\eeq
with the exception of an additive constant,
where $W$ is the probability ({\it Wahrscheinlichkeit}
in the original paper) that the $N$ resonators
have a total energy $U$, and $k$ is one of the 
two constants of nature introduced by Planck
\cite{planck1900}, which is
the Boltzmann constant. Following Boltzmann, Planck
determines the number of complexions which he says is
proportional to $W$. Although, Planck speaks of $W$
as a probability, 
the quantity is in fact understood in formula
(\ref{110}) as the reciprocal of the probability.

Gibbs considered two
types of entropy, given by (\ref{111a}) and (\ref{111b}).
The first form is similar to (\ref{110}) in the sense
that both are proportional to the logarithm of the
number of states of a system with a fixed energy.
Of the two types of entropy, Gibbs opted for the
first. These two types of entropy considered by Gibbs
refer to the microcanonical distribution. He also introduced a form 
of the entropy for systems described by the canonical
distribution. In this case Gibbs
defined entropy as the average of the index of
probability with the sign reversed. As the index of
probability is the logarithm of the probability, 
the Gibbs canonical entropy is
\beq
S_{\rm G} = - k\int \rho \ln\rho dqdp.
\label{140}
\eeq
Although Gibbs speaks of average, in fact the entropy given 
by (\ref{140}) is not an average of a state
function, as $\ln\rho$ is not properly a state function.

It is worth comparing the expression (\ref{140}) with that related
to $H$ given by equation (\ref{133}). If we consider
a system of $N$ particles, the formula (\ref{133})
gives the expression
\beq
S_{\rm B} = - N \int f \ln f d^3rd^3v.
\label{140a}
\eeq
If the particles do not interact, both expression (\ref{140})
and (\ref{140a}) give 
the same value. If the particles interact, $S_{\rm B}$
is distinct from $S_{\rm G}$.

As we have seen above,
the invariance of the second form of the microcanonical
entropy, given (\ref{111b}), was the concern of 
Paul Hertz, who demonstrated the invariance in a paper
published in 1910 \cite{hertz1910}. 
The invariance of the entropy was also the concern of
Einstein. In a paper of 1914, he considered the entropy
of a quantum system with a discretized spectrum of energies
which depended on a parameter \cite{einstein1914}.
He used the Planck formula (\ref{110}),
relating the entropy with the number $Z$ of possible quantum
states, and asked whether the entropy remains valid when the 
states of the system varies under the change of
the parameter. Using the Ehrenfest principle he concluded
that $Z$ is indeed an invariant under the slow parametric action 
and so is the entropy.

\subsection{Work and heat}

The law of conservation of energy tells that
the increase of the energy of a system equals the 
work done on the system. Thermodynamics distinguishes
two types of work. One of them is the heat $Q$, sometimes
called internal work, and the other is the external
work $W$, of simply work. The variation $\Delta U$ of the
energy is thus $\Delta U=Q+W$.
In thermodynamics, the distinction
is provided by the introduction of adiabatic walls.
If a system is enclosed by adiabatic walls, there
is no heat involved and the increase in energy is
the external work. However, we are faced here with
a circular reasoning and another way of distinguish
heat and work is necessary \cite{oliveira2019a}.

The distinction between the two types of work
is obtained by considering the connection
of a system with the environment. One of them is
the ordinary interaction, that we call
{\it dynamic connection}, in which the state
of the system varies by virtue of the connection
of the dynamic variables of the system with
those of the environment. The other is the
{\it parametric connection} in which the state
of the system varies by virtue of the variation
of a parameter. A more precise distinction 
is provided by considering the external forces
acting on the system. In the dynamic connection,
the external forces, depend on the dynamic variables
of the environment but not on the parameters.
In the parametric connection, the external forces
depend on the parameter but not on the dynamic
variables of the environment.

These two types of connection with the environment,
allows us to mechanically distinguish $Q$ and $W$.
If the system is only parametrically connected with the
environment, there is no heat involved, $Q=0$, and the
change of energy of the system is the work $W$ caused
by the variation of the parameter. If the parameter is
held constant and the system is dynamically connected
with the environment, there is no work related to the
variation of the parameter, $W=0$, and the variation in
energy is the work $Q$ related to the dynamic variables,
which is understood as heat.

If the system is connected to the surrounding only
through the parametric action, the resulting process
is thus an adiabatic process, in the thermodynamic sense.
The idea of adiabatic process as a purely mechanical
process is implicit in the works of Clausius and Boltzmann
on the kinetic theory as they considered a thermodynamic
system as a mechanical system. According to Jammer
\cite{jammer1966},
the idea of an adiabatic process as related
with the slow variation of a parameter is to be found
in the works of Helmholtz and Heinrich Hertz,
who tried to identify a parameter of the system
to a cyclic variable.
As we have seen above a precise understanding
of a adiabatic process as the result of a
variation of a parameter was
implicit in the paper of Paul Hertz, who developed
this concept from a hint given by Gibbs.
It appears that it was on this understanding 
that Einstein called the Ehrenfest hypothesis
about parametric invariants the adiabatic
hypothesis, and that Ehrenfest called the
parametric invariant the adiabatic invariant.

If the system is connected to the surrounding only
through the parametric action, the resulting process
is an adiabatic process no
matter if the resulting process is slow or not. The
distinction between a slow and a rapid process is that
in the slow process the system remains in equilibrium
or rather near equilibrium as we are speaking of
a process evolving in time. Processes of this
type are usually called quasi-static processes,
and for that reason sometimes the slow parametric
action is named quasi-static. We think that this
is also inappropriate due to the existence of
quasi-static processes, such as the isothermal
process, which are not adiabatic. Thus we may
say that the slow parametric action results in a
{\it quasi-static adiabatic} process
or a {\it slow adiabatic} process.

\subsection{Equilibrium thermodynamics}

The existence of parametric invariance in classical and
quantum mechanics is the crucial feature that 
allows the construction of a continuous sequence
of equilibrium thermodynamic states when a parameter
is chan\-ged slowly in time, as may happen when
work is done upon a system. This is understood as
follows. If a system in equilibrium
is perturbed during a finite interval of time $\Delta t$, 
it will be found in a nonequilibrium state at
the end of the perturbation, and we have to wait
another finite interval of time $\tau$ for the system
to reach equilibrium again. Suppose that the perturbation
is carried out by the variation of a parameter $\lambda$
which changes by $\Delta\lambda=c\Delta t$ during the interval
$\Delta t$, where $c$ is the rate in which the parameter
is changed. If $c$ is small enough, the adiabatic invariance
guarantees that $\tau$ becomes negligible and the
system remains in equilibrium after the {\it finite} change
$\Delta\lambda$ of the parameter.

In accordance with the understanding  
concerning heat and work,
we write the variation of the energy during a certain 
small interval of time when the parameters $\lambda_i$
is slowly varying in time as
\beq
dU = dQ + dW,
\label{102c}
\eeq
where $dW$ is the work performed by the system.
As the parameter are slowly varying in time,
the resulting process is adiabatic, $dQ=0$
and $dW=dU$. Using the expression (\ref{102}) 
for $dU$, we write
\beq
dW = - \sum_i F_i d\lambda,
\label{102a}
\eeq
where
\beq
F_i =  - \overline{\frac{\partial H}{\partial\lambda}},
\eeq
and ${\cal H}$ is the Hamiltonian of the system.
The time average is in turn replaced by the average 
\beq
F_i =  - \la\frac{\partial H}{\partial\lambda}\ra,
\label{102b}
\eeq
over a Gibbs probability distribution $\rho$.
The energy $U$ is understood as the average 
$U=\la{\cal H}\ra$.

If we consider a slow variation of the parameter,
this process will corresponds to lines that lay
down on a surface on the space spanned
by the parameters $\lambda_i$ and by the energy $E$.
Along any lines there is no exchange of heat
and the parametric invariant $\Phi$ is constant.
That is, each surface
is characterized by a certain value of $\Phi$. 
The energy can then be considered as function of
$\Phi$ and of the parameters, so that
\beq
dU = A d\Phi - \sum_i F_i d\lambda_i,
\eeq
which compared to (\ref{102c}) tell us that
\beq
dQ = A d\Phi.
\label{102d}
\eeq

The relation (\ref{102d}) is in accordance with 
the Clausius relation $dQ=TdS$. However, 
it does not mean necessarily that the entropy $S$
is equal or proportional do $\Phi$. It says that
$S$ is a function of $\Phi$. 
To determine the specific
function, we must introduce a condition. 
We may suppose that $S$ scales as the size of the system.
Taking into account that $\Phi$ scales with the size of the system
to the $n$-th power, where $n$ is the number of degrees of
freedom, we may place $\ln\Phi$ as proportional
to $S$ and write
\beq
S = k\ln\Phi,
\label{28}
\eeq
where $k$ is a constant with the same physical dimension of
the entropy. 

Having established the relation between the entropy $S$ and
the invariant $\Phi$,
we may determine the temperature $T$ by its thermodynamic
definition $1/T=\partial S/\partial E$. In the case of a
system described by the Gibbs microcanonical distribution,
$\Phi$ is given by (\ref{22a}), and  we find
\beq
\frac1{kT} = \frac{\Omega}{\Phi},
\eeq 
where $\Omega$ is given by (\ref{22b})
and is related to $\Phi$ by $\Omega=\partial \Phi/\partial E$.

\subsection{Canonical distribution}

If a system is in contact with
the environment only through a parametric connection,
the Hamiltonian ${\cal H}$ is a function
of the internal variables, which are the dynamic
variables $q$ and $p$ of the system. 
The external forces depend on the parameters and
on the internal variables but not
on the external variables, which are the variables
of the environment.
In this case the energy of the system is the
average of ${\cal H}$ which is a function of the
internal variables.

Now let us consider the case where the environment is
subject to the dynamic connection. In this case the
external forces are functions of both the internal
and external variables and the energy of the system
will have a term that is related to the external
variables. If the system is nearly equilibrium the
treatment given by the statistical mechanics
is to replace the dynamic description by 
a description in terms of a probabilistic description
given for example by the Gibbs canonical description.
In this description the probability distribution 
is a function of a Hamiltonian which is
understood as the Hamiltonian ${\cal H}$ that one obtains by 
removing the terms describing the dynamic interaction
with the environment. This does not mean that
the interaction has been neglected. In fact, it is taken
into account through the probability distribution itself,
such as the Gibbs canonical distribution which 
describes the interaction with a heat reservoir
at a given temperature. Therefore, the energy $U$ that enters
equation (\ref{102c}) is the average of ${\cal H}$
which depends only on the internal dynamic variables.

In the present case of the dynamic connection,
in which the system is described by the Gibbs canonical
distribution the parametric invariant is $\Psi$
given by equation (\ref{45a}),
\beq
\Psi = e^{\beta U} Z,
\eeq
Again the usual choice is a logarithm relation
\beq
S = k \ln\Psi = k\beta U + k\ln Z,
\label{130}
\eeq
which gives an entropy that increases with the size of the
system. Deriving this relation with respect to 
$\beta$ and using relation (\ref{127}), and comparing
with the Clausius relation $dU=TdS$, one finds 
$\partial S/\partial\beta = k\beta \partial U/\partial\beta$,
which gives the relation $\beta=1/kT$ between the parameter
$\beta$ and $T$.

What is the relation between the invariant $\Psi$
and the Gibbs expression for the entropy 
\beq
S_{\rm G} = - k \int \rho \ln \rho dqdp.
\eeq 
If we replace the distribution (\ref{125}) in this
equation we find that it equals $k \ln \Psi$, that is,
it is an invariant and coincides with the entropy
(\ref{130}) that scales with the size of the system.

\section{Conclusion}

We have argued that the slow variation of a parameter
of a system results in a slow adiabatic process.
Thus a parametric invariant such as the Hertz
invariant is an adiabatic invariant. As the entropy
is a thermodynamic variable that characterizes
a reversible adiabatic process, 
understood as slow adiabatic process, it is
constant along this process and can be understood
as a parametric invariant. As the entropy grows with
the size of the system, it is defined as the logarithm
of the Hertz invariant. 

The Hertz invariant $\Phi$ is the volume in phase space
of the region enclosed by the surface of constant
energy. It remains constant when a parameter is
slowly changed with time. We have extended this result
to the case where the number of particles varies
with time with the conclusion that $\Phi^*=\Phi/N!$ is
the invariant. We have also considered the quantum version
of the Hertz invariant for the case of the variation
of a parameter as well as the variation of the number
of particles. In this last case, the classical limit
results in the quantity $\Phi^*=\Phi/N!$

The parametric invariance allows the distinction
between heat and work. This is provided by considering
the two types of connections of the system with
the environment that we have called
dynamical and parametric connection. If only the
second is present and the parameter is varied,
the resulting process is adiabatic. If in addition
the variation of the parameter is slow, the 
entropy characterizes this process and can thus
be associated to the parametric invariant.


\end{document}